\documentclass[aps,groupedaddress,superscriptaddress,showpacs]{revtex4-1}
\usepackage{eurosym}
\usepackage{amsmath}
\usepackage{fixmath}
\usepackage{graphicx}
\usepackage{wrapfig}
\usepackage[toc,page]{appendix}
\usepackage{caption}
\usepackage{hyperref}
\usepackage[capitalise]{cleveref}
\usepackage{subcaption}
\usepackage{amsfonts}
\usepackage{xcolor}

\graphicspath{{images/}}

\newcommand{\be}{\begin{equation}}
\newcommand{\ee}{\end{equation}}
\newcommand{\eq}[1]{Eq.~(\ref{#1})}

\DeclareMathOperator{\arcosh}{arcosh}

\begin{document}

\title{Dissipation-driven formation of entangled dark states in strongly-coupled inhomogeneous many-qubit systems in solid-state nanocavities}

\author{ Mikhail Tokman}
\affiliation{ }
\author{Alex Behne}
\affiliation{Department of Physics and Astronomy, Texas A\&M University, College Station, TX, 77843 USA}
\author{Brandon Torres}
\affiliation{Department of Physics and Astronomy, Texas A\&M University, College Station, TX, 77843 USA}
\author{ Maria Erukhimova}
\affiliation{ }
\author{Yongrui Wang}
\affiliation{Department of Physics and Astronomy, Texas A\&M University, College Station, TX, 77843 USA}
\author{Alexey Belyanin}
\affiliation{Department of Physics and Astronomy, Texas A\&M University, College Station, TX, 77843 USA}

\begin{abstract}

We study quantum dynamics of many-qubit systems strongly coupled to a quantized electromagnetic cavity field in the presence of decoherence and dissipation for both fermions and cavity photons, and taking into account the varying coupling strength of different qubits to the cavity field and the spread of their transition frequencies.  Compact analytic solutions for time-dependent quantum state amplitudes and observables are derived for a broad class of open quantum systems in Lindblad approximation with the use  of the stochastic Schroedinger equation approach. We show that depending on the initial quantum state preparation, an ensemble of qubits can evolve into a rich variety of many-qubit entangled states with destructive or constructive interference between the qubits. In particular, when only a small fraction of qubits is initially excited, the dissipation in a cavity will inevitably drive the system into robust dark states that are completely decoupled from the cavity and live much longer than the decay time of the cavity field. We also determine the conditions under which coherent coupling to the quantized cavity field overcomes the dephasing caused by a spread of transition frequencies in multi-qubit systems and leads to the formation of a decoupled dark state.

\end{abstract}

\date{\today }

\maketitle

\section{Introduction}

Solid-state cavity quantum electrodynamics (QED) attracted much interest as a promising platform
for quantum information and quantum sensing systems; see, e.g., \cite{thorma2015, lodahl2015, degen2017,  dovzhenko2018, bitton2019} for recent reviews. A typical scenario
involves an ensemble of quantum emitters (ideally, two-level systems) such
as quantum dots, defects in crystals, or molecules, strongly coupled to a
quantized electromagnetic (EM) field in a dielectric or plasmonic nanocavity. We will call these quantum
emitters qubits for brevity, although the logical qubits forming the gates may involve many
two-level systems as well as photonic or mixed degrees of freedom. Several
or many qubits are required for most applications. Although direct
near-field coupling among qubits is possible and desired for some gating protocols, in the nanophotonics context 
such coupling would require deterministic placement of qubits with sub-nm
accuracy, which is challenging. A simpler scenario which still permits various ways of quantum state manipulation is the one in which the
qubits are coupled only through the common cavity field. This is the situation considered in this paper. 

The problem of $N$ qubits strongly coupled to a quantized cavity mode has of course been considered many times, starting from the seminal Tavis-Cummings paper \cite{tavis1968}. Inherent in most of these studies is the assumption of identical qubits coupled to the field with identical coupling strengths. This makes the system invariant to permutations and allows one to drastically reduce the number of degrees of freedom and related computational effort; see, e.g., the recent work \cite{shammah2018} (and references therein) where an efficient numerical solver was proposed to solve the N-qubit master equation in Lindblad approximation.  A clean case of the Tavis-Cummings quantum dynamics in the strong-coupling regime was recently observed for a large ensemble of donor spins in a microwave cavity \cite{rose2017}, taking advantage of their minimal inhomogeneous broadening and uniform spin-cavity coupling. 

In the solid-state nanocavity context, the cavity field is strongly nonuniform, especially in plasmonic nanocavities where it varies on a nanometer scale. This makes the qubit-cavity coupling strength strongly variable from qubit to qubit. Moreover, for many popular quantum emitters, such as quantum dots, optically active point defects, excitons in semiconductor nanostructures etc., the spread of transition frequencies exceeds homogeneous linewidth, making inhomogeneous broadening the dominant source of dephasing. Any of these factors break permutation symmetry and increase the complexity of the problem, making it difficult to solve even numerically for large $N$. As a result, the problems with dissimilar quantum emitters are usually analyzed for few qubits, and even then numerical treatment of the Lindblad master equation is required, e.g., \cite{laucht2010,gray2015,gray2016}.

 Here we are able to drastically simplify the analysis and obtain analytic or semi-analytic solutions for quantum dynamics of $N$ strongly coupled dissimilar qubits or multilevel fermionic systems in the presence of decoherence and dissipation for both fermions and cavity photons. This progress is made possible by applying a modified version of the stochastic Schr\"{o}dinger equation (SSE) formalism. The idea of adding Langevin noise to the Schr\"{o}dinger equation is nothing new; see, e.g., \cite{zoller1997,Plenio1998, gisin1992,diosi1998, cohen1993, molmer1993,gisin1992-2}. This approach is typically used for numerical Monte-Carlo simulations. We recently developed a version of SSE suitable for analytic solutions of open strongly-coupled cavity QED problems  \cite{tokman2020, chen2021} and, as we show here, it is quite useful in analysis of nonuniform and inhomogeneously broadened many-qubit systems.  Note that the most popular stochastic approach, namely the Heisenberg-Langevin formalism faces challenges when dealing with nonperturbative dynamics of strongly coupled systems as it leads to nonlinear operator-valued equations. 
 
The possibility of analytic treatment comes with limitations. We use the rotating wave approximation (RWA) throughout the paper, which means that all frequency scales such as Rabi frequencies and detunings are much smaller than the cavity mode frequency and the optical transition frequencies in the qubits. Beyond the RWA, general analytic solutions become impossible in the fully quantized and nonlinear (nonperturbative) regime of light-matter interaction between fermionic and bosonic fields. Note that some recent experiments with low-frequency transitions (e.g. microwave or terahertz) in strongly coupled multi-electron systems went beyond the RWA into the ultra strong coupling regime; e.g., \cite{todorov2010, forndiaz2017, kono2019}.  

Since we have a complete analytic solution for many cases, we can calculate any observables. In this paper we cherry-picked some of the more interesting examples related to the dissipation-driven formation of highly entangled dark states that are decoupled from the cavity field. The ability to generate and control such states is a problem of great practical importance for the rapidly developing field of plasmonic nanocavity QED \cite{chikkaraddy2016,benz2016, park2016, pelton2018, gross2018, park2019},  where the dissipation of
a cavity mode is much faster than the relaxation in quantum emitters. 

The dissipation-driven formation of entangled bright and dark states in ensembles of quantum emitters has been studied extensively in the context of the Dicke model of superradiance \cite{dicke1954,gross1982}; see, for example,  \cite{schneider2002, gonzales2013, scully2015,  kirton2017, wolfe2014, shammah2017, gegg2018} and references therein.  The typical bad-cavity or no-cavity regime of Dicke superradiance is in a sense opposite to the regime of strong-coupling dynamics, although extended samples can still demonstrate complex oscillatory quasi-chaotic propagation effects  \cite{belyanin1998, cong2016}.  Furthermore, the analysis of Dicke superradiance is usually based on collective angular momentum operators which imply permutation symmetry, although some studies do include inhomogeneous ensembles of quantum emitters \cite{temnov2005}. Our analysis points out certain similarities between the entangled bright or dark states in the Dicke superradiance problem vs. strongly coupled systems, as well as important peculiarities of strongly coupled dynamics. 

Perhaps one of the greatest advantages of our SSE-based approach is transparent description of entanglement which is, in most cases, obvious from the explicit analytic form of the state vector. In contrast, the characterization of entanglement within the master equation formalism is a separate problem, since convenient universal figures of merit exist only for simplest systems; see, e.g., \cite{guhne2009} for a review and \cite{gray2015,gray2016} for recent studies of entanglement in strongly coupled nanocavity QED systems based on the master equation. A variety of entangled state control scenarios were studied for microwave cavities within superconducting circuit QED; see \cite{tureci2016} and references therein. 

In Sec.~II we introduce the Hamiltonian and general classification of quantum states for $N$ two-level qubits strongly coupled to a quantized cavity mode, including the spread of transition frequencies and coupling strengths. Section III contains a short outline of the description of dissipation, dephasing, and noise in strongly-coupled systems based on the stochastic equation of evolution, which we introduced in more detail elsewhere \cite{tokman2020, chen2021}. Section IV derives general analytic solutions for quantum dynamics in the case of single-photon excitation energies and provides examples illustrating dissipation-driven formation  of entangled dark states decoupled from the cavity field as well as the corresponding emission spectra. In Sec.~V we generalize the treatment to an ensemble of qubits with a large spread of transition frequencies and to band-to-band transitions in multi-level electron systems. This introduces additional
effective decoherence and quasi-chaotic dynamics of individual qubits. Surprisingly, even in this case there still exist entangled dark states as long as the collective Rabi frequency exceeds the total spread of transition frequencies. In Section VI we provide a general formalism and classification of bright and dark states for arbitrary $M$-photon excitations in dissipative strongly coupled $N$-qubit systems and illustrate this formalism with analytic results for small values of $M$ and $N$ and numerical examples.   Appendix A derives some useful analytic formulas for the spatial field distribution in the practically important case of a nanocavity formed by a metallic nanotip or a nanoparticle over a metallic substrate, which has been used in a variety of recent experiments.  Appendix B derives approximate analytic results for quantum dynamics of inhomogeneously broadened ensembles of qubits.

Since most results in this paper are in the analytic form and the plots are normalized, here, we list typical values of the parameters in experimental solid-state nanophotonic systems which determine the strength of light-matter coupling and relaxation rates. To determine if the strong coupling regime and quantum entanglement in the electron-photon system can be achieved,  one has to compare the relaxation rates with the characteristic coupling strength between the cavity field and the qubits. Whereas for identical qubits in a uniform field such a coupling parameter is simply the Rabi frequency, the corresponding ``figures of merit'' become nontrivial in a nanocavity where individual quantum emitters experience a strongly nonuniform field distribution of a cavity mode; see Sec.~IV. The strong coupling criterion becomes even more complicated with many different regimes possible if the qubits have a broad distribution of transition frequencies  which lead to additional dephasing as shown in Sec.~V. 

The parameters mentioned below were obtained from recent experimental reports and reviews such as \cite{thorma2015, lodahl2015, degen2017,  dovzhenko2018, bitton2019, todorov2010, forndiaz2017, kono2019, chikkaraddy2016,benz2016, park2016, pelton2018, gross2018, park2019, lukin2016,deppe,reithmaier,sercel2019, fieramosca2019} . This paper does not attempt to provide a comprehensive overview of the rapidly growing literature on the subject.

In electron-based quantum emitters, the largest oscillator strengths in the visible/near-infrared range have been observed for excitons in organic molecules, followed by perovskites and more conventional inorganic semiconductor quantum dots. The typical variation of the dipole matrix element of the optical transition which enters the Rabi frequency is from tens of nm to a few Angstroms (in units of the electron charge). Within the same kind of system, the dipole moment grows with increasing wavelength. The relaxation times are strongly temperature and material quality dependent, varying from tens or hundreds of ps for electric-dipole allowed interband transitions in quantum dots at 4 K to the tens and hundreds of $\mu$s for spin qubits based on defects in semiconductors and diamond at mK temperatures. Dipole-forbidden optical transitions, e.g., dark excitons in quantum dots, can have lifetimes of up to milliseconds, but the coupling to light of these transitions could be too weak for them to reach the strong coupling regime.  At room temperature, the relaxation times for the optical transitions are in the ps range or shorter.

Photon decay times are longest for dielectric microcavities: photonic crystal cavities, nanopillars, distributed Bragg reflector mirrors, microdisk whispering gallery mode cavities, etc. Their quality factors are typically between $10^3 - 10^7$, corresponding to photon lifetimes from sub-ns to $\mu$s. However, the field localization in the dielectric cavities is diffraction-limited, which limits the attainable single-qubit vacuum Rabi frequency values to hundreds of $\mu$eV \cite{deppe,reithmaier}. The effective decay rate of the eigenstates (e.g. exciton-polaritons)  in dielectric cavity QED systems is typically limited by relaxation in fermion quantum emitter subsystem. 

In plasmonic cavities, field localization on a nm and even sub-nm scale has been achieved, but the photon decay time is in the tens of fs range and therefore, the photon losses dominate the overall decoherence rate. Still, when it comes to strong coupling at room temperature to a single quantum emitter such as a single molecule or a quantum dot, the approach utilizing plasmonic nanocavities has seen more successful so far. In these systems, single-emitter Rabi splitting on the order of 100--200 meV has been observed \cite{chikkaraddy2016,benz2016, park2016, pelton2018, gross2018, park2019}.

\section{ N qubits in a nonuniform nanocavity field: The model}

We have in mind a typical scenario with a few or many qubits located in a nonuniform field of a solid-state nanocavity formed, e.g., by a nanotip and a metallic substrate as sketched in Fig.~1a or a graphene nanostructure supporting surface plasmon-polariton modes as in Fig.~1b (e.g., \cite{manj2012,chen2017}). \\


\begin{figure*}[htb]
\centering

\begin{subfigure}[b]{0.4\textwidth}
\includegraphics[width=\textwidth]{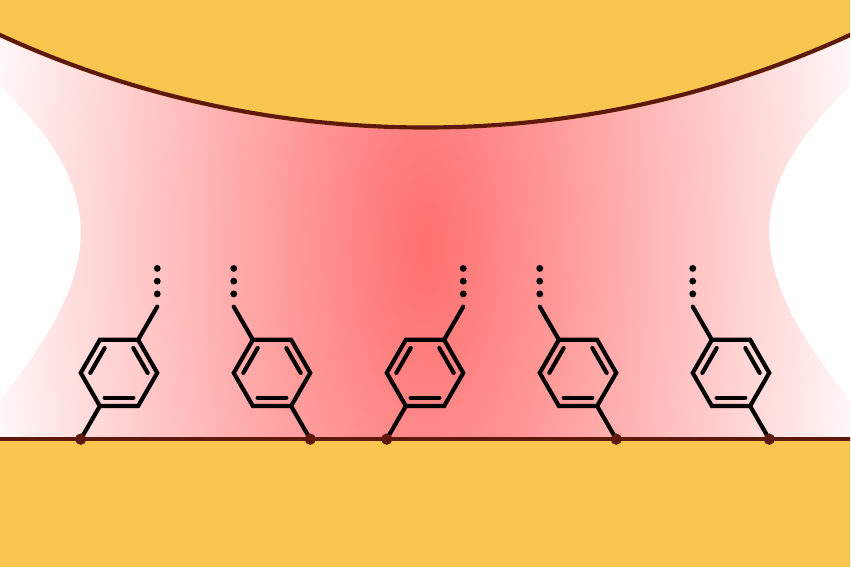}
\caption{ }
\label{fig1}
\end{subfigure}
\hfill

\begin{subfigure}[b]{0.5\textwidth}
\centering
\includegraphics[width=\textwidth]{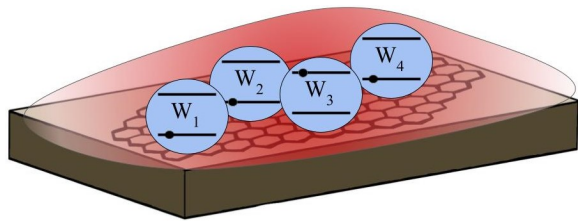}
\caption{}
\label{fig1}
\end{subfigure}

\caption { An ensemble of quantum emitters (e.g. quantum dots or  molecules) in a nanocavity consisting of (a) 
a metallic nanoparticle or nanotip of the scanning probe and a metallic substrate, or (b) a graphene nanopatch supporting a surface plasmon-polariton mode.  }

\end{figure*}


We begin by introducing the Hamiltonian and defining the variables for a system of $N$  two-level systems with
states $\left\vert 0_{j}\right\rangle $ and $\left\vert 1_{j}\right\rangle $%
, where $j=1,...N$, with energy levels $0$ and $W_j$. We introduce fermionic
operators of annihilation and creation of an excited state $\left\vert
1_{j}\right\rangle $,
\begin{equation}
\hat{\sigma}_{j}=\left\vert 0_{j}\right\rangle \left\langle 1_{j}\right\vert
,\ \ \ \hat{\sigma}_{j}^{\dagger }=\left\vert 1_{j}\right\rangle
\left\langle 0_{j}\right\vert ,  \label{sigma j}
\end{equation}%
the dipole moment operator, 
\begin{equation}
\mathbf{\hat{d}}=\sum_{j=1}^{N} \left( \mathbf{d_j} \hat{\sigma}_{j}^{\dagger }+ \mathbf{d^*} 
\hat{\sigma}_{j}\right),  \label{dmo}
\end{equation}
and the Hamiltonian for all qubits,
\begin{equation}
\hat{H}_{a}= \sum_{j=1}^{N} W_j \hat{\sigma}_{j}^{\dagger }\hat{\sigma}_{j}.
\label{Ha}
\end{equation}%
Here $\mathbf{d_j} = \left\langle 1_{j}\right\vert \mathbf{\hat{d}}\left\vert
0_{j}\right\rangle $. Our $N$-qubit system interacts
with a single-mode field 
\begin{equation}
\mathbf{\hat{E}}=\mathbf{E}(\mathbf{r}) \hat{c}+\mathbf{E}^{\ast
}(\mathbf{r}) \hat{c}^{\dagger },  \label{E}
\end{equation}%
where $\hat{c}$ and $\hat{c}^{\dagger }$ are standard annihilation and
creation operators for bosonic Fock states. 

The function $\mathbf{E}(\mathbf{r}) $ is the spatial
structure of the electric field in a cavity. It is normalized as in \cite{Tokman2016}
\begin{equation}
\int_{V}\frac{\partial \left[ \omega ^{2}\varepsilon \left( \omega ,\mathbf{r%
}\right) \right] }{\omega \partial \omega }\mathbf{E}^{\ast }(\mathbf{r}) \mathbf{E}(\mathbf{r}) d^{3}r=4\pi \hbar \omega
\label{nc-const}
\end{equation}%
to preserve the standard form of the field Hamiltonian,
\begin{equation}
\hat{H}_{em}=\hbar \omega \left( \hat{c}^{\dagger }\hat{c}+\frac{1}{2}\right). 
\label{HEM}
\end{equation}%

The relation between the modal frequency $\omega $ and the function $\mathbf{%
E}(\mathbf{r}) $ can be found by solving the classical electrodynamics boundary-value problem corresponding to the cavity in question. Here $V$ is a quantization volume
and $\varepsilon \left( \omega ,\mathbf{r}\right) $ is the dielectric
function of a dispersive medium that fills the cavity.

The total Hamiltonian after adding the electric-dipole interaction with the field within the RWA is
\begin{equation} 
\hat{H}=\hbar \omega \left( \hat{c}^{\dagger }\hat{c}+\frac{1}{2}\right)
+\sum_{j=1}^{N} W_j\hat{\sigma}_{j}^{\dagger }\hat{\sigma}_{j}   
-\hbar \sum_{j=1}^{N} \left( \Omega_{Rj} \hat{\sigma}_{j}^{\dagger }\hat{c} + h.c. \right),  
\label{RWAH}
\end{equation} 
where $\Omega_{Rj} = \frac{\mathbf{d_j}\cdot \mathbf{E}\left( \mathbf{r_j}\right)}{\hbar}$ is the Rabi frequency for the $j$th qubit located at the position $ \mathbf{r_j}$ in the cavity. 
Note that this model includes a spread of the transition energies of the qubits $W_j$, and the variation of the cavity EM field depending on the position of each qubit, which is essential for any nanocavity. Therefore, the model loses permutation symmetry which was used to drastically simplify the analysis in \cite{tavis1968, shammah2018}.  Nevertheless, as we show below, a significant reduction in the dimensionality of the problem is possible in our case too. 

Hereafter, we will use the Hamiltonian in the interaction picture:
\begin{equation}
\hat{H}_{int} =    
-\hbar \sum_{j=1}^{N} \left( \Omega_{Rj} \hat{\sigma}_{j}^{\dagger }\hat{c} e^{i \Delta_j t} + h.c. 
\right),  
 \label{hint}
\end{equation}
where $\Delta_j$ = $\frac{W_j}{\hbar} - \omega$. 


For an arbitrary quantum state of the N-qubit system coupled to a cavity mode, the state vector can be expanded over all possible combinations of subsystems as 
\begin{equation}
\label{a1} 
\Psi = \sum_{n=0}^{\infty} \sum_{p = 0}^N \sum_{\alpha_p = 1}^{\mathcal{C}_N^p} C_{n p \alpha_p} | n \rangle |p, \alpha_p \rangle , 
\end{equation}
where $| n \rangle$ is a Fock state of the boson (EM) field and $|p, \alpha_p \rangle$ is a fermion state. Here, the index $\alpha_p$ denotes {\it different} subsets of $p$ elements out of a set of $j = 1,2,\dots N$, which correspond to the excitation of $p$ qubits out of $N$. The total number of such subsets is determined by the binomial coefficient $\mathcal{C}_N^p = \frac{N!}{p! (N-p)!}$. The state $|p, \alpha_p \rangle$ can be written as 
\begin{equation}
|p, \alpha_p \rangle = \left( \prod_{j_p \in \alpha_p} \left\vert \hat{\sigma}_{j_p}^{\dagger} \right\rangle \right) |0_{qub} \rangle, \nonumber
\end{equation}
where $j_p \in \alpha_p$ are qubit numbers belonging to the subset marked by index $\alpha_p$ and  
\begin{equation}
\label{a3} 
|0_{qub} \rangle = \prod_{j=1}^{N} |0_j \rangle = |0,\alpha_0 \rangle.
\end{equation} 

As a reminder, when the coefficients $C_{n p \alpha_p}$ are calculated using the Hamiltonian (\ref{hint}) then the operators used to calculate the observables should be transformed in the same way the Hamiltonian (\ref{RWAH}) was transformed into the interaction picture, Eq.~(\ref{hint}), namely $\hat{\sigma}_{j} \rightarrow \hat{\sigma}_{j} e^{-i\frac{W_j}{\hbar} t}$ and $\hat{c} \rightarrow \hat{c} e^{-i\omega t}$.

Similar to the case of identical qubits and identical field strength at the locations of each qubits \cite{tavis1968}, the Schr\"{o}dinger equation with the Hamiltonian (\ref{hint}) leads to a set of linear equations for the probability amplitudes  
$C_{n p \alpha_p}$ which can be split into independent blocks corresponding to the condition
\begin{equation}
\label{a5} 
n+p = M = {\rm const.}
\end{equation} 
The dimension of the Hilbert space within each independent block is $\sum_{p=0}^{{\rm min}[M,N]} \mathcal{C}_N^p$; for $M \geq N$ it is equal to $\sum_{p = 0}^{N}  \mathcal{C}_N^p = 2^N$. Further reduction of the dimensionality of the problem would require identical values of the $W_j$ and $\Omega_{Rj}$ in which case all states including initial conditions have exact permutation symmetry and one could sum over all states corresponding to various combinations $\alpha_p$ made of $p$ excited atoms; see \cite{shammah2018} and the discussion in Section VI. 

As we discuss in Section VI, in the presence of dissipation and noise, the noise source terms couple the groups with different values of $M$. However, in the strong coupling regime, such noise-induced coupling scales as a small ratio of dissipation rates to the Rabi frequency  and therefore can be included perturbatively. A similar perturbative approach has been developed for nonlinear strong coupling of electron-photon-phonon systems \cite{parametric}.

In practice, 
 the generation of nonclassical {\it multiphoton} states is still a tremendous experimental challenge. The multiphoton fields used in experiments are typically coherent classical laser pulses. In this case, the problem of excitation of qubits by a classical laser field is drastically simplified and is outside the scope of this paper. Therefore, in the main part of this paper we choose initial conditions corresponding to single-photon excitation energies.  Single-photon sources of quantum light are readily available and can be used for initialization of both single- and many-qubit states with a single-photon excitation energy.  These are widely used in quantum information applications, including, for example, Bell states and their generalizations to many-qubit systems; see, e.g., \cite{langford2011,reitz2022}. With single-photon excitations as initial conditions, the states that can be reached as a result of evolution of the system 
 include the ground state $\left\vert 0\right\rangle
\Pi_{j=1}^{N}\left\vert 0_{j}\right\rangle $ and the states with energies
close to the single-photon energy:
\begin{equation}
\Psi =C_{00}\left\vert 0\right\rangle \Pi _{j=1}^{N}\left\vert
0_{j}\right\rangle +C_{10}\left\vert 1\right\rangle \Pi _{j=1}^{N}\left\vert
0_{j}\right\rangle +\sum_{j=1}^{N}C_{0j}\left\vert 0\right\rangle \left\vert
1_{j}\right\rangle \Pi _{m\neq j}^{N}\left\vert 0_{m}\right\rangle.
\label{Psi}
\end{equation}

A nontrivial general result of this work is analytic or semi-analytic solutions for the quantum dynamics of these states for an arbitrary number $N$ of qubits, arbitrary cavity field nonuniformity, arbitrary distribution of transition frequencies and Rabi frequencies (as long as the RWA is still valid), and in the presence of dissipation and noise for both matter and the field.  Since we have analytic results for all probability amplitudes and therefore all possible observables, in this paper we cherry-pick some of the most interesting examples such as the spontaneous formation of long-lived entangled dark states or interplay between inhomogeneous broadening of qubit frequencies and coherence in the system. 

Arbitrary multiphoton excitations with fully quantized multiphoton states can be treated within the same formalism but involve far more painful algebra. We provide general classification of quantum dynamics for arbitrary $M$-photon excitations in terms of bright and dark states in Section VI, as well as some analytic and numerical examples for small $M$.

\section{Description of dissipation and noise using stochastic equations of evolution} 

A standard way to include the effects of dissipation is based on the master
equation for the density matrix $\hat{\rho}$ of the system \cite{blum},
\begin{equation}
\frac{d}{dt}\hat{\rho}=-\frac{i}{\hbar }\left[ \hat{H},\hat{\rho}\right] +%
\hat{L}(\hat{\rho}),  \label{meq1}
\end{equation}%
where $\hat{L}(\hat{\rho})$ is the relaxation operator. If there are $S$
states in a given basis $|\alpha \rangle $, Eq.~(\ref{meq1}) corresponds to $%
\frac{1}{2}S(S+1)$ equations for the matrix elements $\rho _{\alpha \beta
}=\rho _{\beta \alpha }^{\ast }$. The number of equations that need to be
solved can be reduced to $S$ via the method of the stochastic equation of
evolution for the state vector 
 \cite{zoller1997, Plenio1998, gisin1992,  diosi1998, cohen1993, molmer1993, gisin1992-2,   tokman2020,chen2021}. This becomes
possible if the structure of the relaxation operator permits representing
the right-hand side of Eq.~(\ref{meq1}) in the form
\begin{equation}
-\frac{i}{\hbar }\left[ \hat{H},\hat{\rho}\right] +\hat{L}(\hat{\rho})=-%
\frac{i}{\hbar }(\hat{H}_{eff}\hat{\rho}-\hat{\rho}\hat{H}_{eff}^{\dagger
})+\delta \hat{L}(\hat{\rho}),  \label{meq2}
\end{equation}%
where $\hat{H}_{eff}=\hat{H}+\hat{H}^{\left( ah\right) }$ is an effective
non-Hermitian Hamiltonian. 

Within the Markovian models of relaxation, the stochastic equation for the
state vector takes the form
\begin{equation}
\frac{d}{dt}\left\vert \Psi \right\rangle =-\frac{i}{\hbar }\hat{H}%
_{eff}\left\vert \Psi \right\rangle -\frac{i}{\hbar }\left\vert \mathfrak{R}%
\right\rangle .  \label{S eq}
\end{equation}%
In Eq.~(\ref{S eq}) the vector $\left\vert \mathfrak{R}\right\rangle $ is a
stochastic Langevin source with the following statistical properties:
\begin{equation}
\overline{\left\vert \mathfrak{R}\right\rangle }=0,\ \ \ \overline{\mathfrak{%
R}_{\alpha }\left( t^{\prime }\right) \mathfrak{R}_{\beta }^{\ast }\left(
t^{\prime \prime }\right) }=\hbar ^{2}\delta \left( t^{\prime }-t^{\prime
\prime }\right) D_{\alpha \beta },\ \ \ D_{\alpha \beta }=\left\langle
\alpha \right\vert \delta \hat{L}(\hat{\rho})\left\vert \beta \right\rangle
_{\hat{\rho}\Longrightarrow \overline{\left\vert \Psi \right\rangle
\left\langle \Psi \right\vert }};  \label{statistical properties}
\end{equation}
the overbar $\overline{\left( \cdots \right) }$ means averaging over the
noise statistics, $\mathfrak{R}_{\alpha }=\left\langle \alpha \right\vert
\left. \mathfrak{R}\right\rangle $ . The dyadics $\overline{C_{\alpha
}C_{\beta }^{\ast }}$ in Eqs.~(\ref{S eq}),(\ref{statistical properties})
where $C_{\alpha }=\left\langle \alpha \right\vert \left. \Psi \right\rangle
$ correspond to the density matrix elements $\rho _{\alpha \beta }$ in the
master equation (see the proof in \cite{tokman2020}).

The observables in the method of the stochastic equation are determined by
\begin{equation*}
g=\overline{\left\langle \Psi \right\vert \hat{g}\left\vert \Psi
\right\rangle },
\end{equation*}%
where $\hat{g}$ is an operator corresponding to the physical quantity $g$.
This definition differs from a standard one by an additional averaging over
the noise statistics. The choice of operators $\hat{H}^{\left( ah\right) }$
and correlators $D_{\alpha \beta }$ should ensure the conservation of the
norm of the stochastic vector, $\overline{\left\langle \Psi \right\vert
\left. \Psi \right\rangle }=1$, and bring the system to a physically
reasonable steady state in the absence of external perturbation.

Another widely used method to include the effects of dissipation in quantum
optics is the Heisenberg--Langevin approach \cite{Scully1997,Gardiner2004}.
However, when applied to the dynamics of strongly coupled systems, the
Heisenberg equations become nonlinear (see, e.g., \cite{Scully1997}),
whereas the stochastic equation for the state vector, Eq.~(\ref{S eq}), is
always linear, which is an important advantage of this method.

The representation of the type shown in Eq.~(\ref{meq2}) is possible, in
particular, for the Lindblad relaxation operator. Here we will use the
Lindbladian $\hat{L}(\hat{\rho})$ in the case of independent dissipative
reservoirs for the field and qubits and at zero temperature: 
\begin{equation}
L(\hat{\rho})=-\Sigma _{j}[\frac{\gamma _{j}}{2}(\hat{\sigma}_{j}^{\dagger }%
\hat{\sigma}_{j}\hat{\rho}+\hat{\rho}\hat{\sigma}_{j}^{\dagger }\hat{\sigma}%
_{j}-2\hat{\sigma}_{j}\hat{\rho}\hat{\sigma}_{j}^{\dagger })]-\frac{\mu }{2}(%
\hat{c}^{\dagger }\hat{c}\hat{\rho}+\hat{\rho}\hat{c}^{\dagger }\hat{c}-2%
\hat{c}\hat{\rho}\hat{c}^{\dagger }),  \label{Lindbladian at 0 T}
\end{equation}%
which gives%
\begin{equation}
\hat{H}_{eff}=\hat{H}-i\hbar \frac{1}{2}\left( \sum_{j}\gamma _{j}\hat{\sigma%
}_{j}^{\dagger }\hat{\sigma}_{j}+\mu \hat{c}^{\dagger }\hat{c}\right) ,
\label{efh}
\end{equation}%
\begin{equation}
\delta \hat{L}(\hat{\rho})=\sum_{j}\gamma _{j}\hat{\sigma}_{j}\hat{\rho}\hat{%
\sigma}_{j}^{\dagger }+\mu \hat{c}\hat{\rho}\hat{c}^{\dagger }.
\label{delta LO}
\end{equation}%
Here the relaxation constants $\mu $ and $\gamma _{i}$ are determined by the
cavity Q-factor and inelastic relaxation of the qubits, respectively. The Q-factor is determined by adding up diffraction and Ohmic losses in a cavity; e.g., \cite{tokman2018,tokman2019}. 
Elastic relaxation processes (pure dephasing) are included later in this section. The case of
arbitrary temperatures is considered in \cite{tokman2020}. Note that for a qubit with the
transition in the visible or near-IR range, even a room-temperature reservoir
is effectively at zero temperature.

Introducing state vectors $\Psi $ of the type given in Eq.~(\ref{Psi}), 
we obtain a set of stochastic equations for the amplitudes,
\begin{equation}
\dot{C}_{00}+ \gamma _{00} C_{00}=-\frac{i}{\hbar }\mathfrak{R}_{00},
 \label{C00 dotted}
\end{equation}%
\begin{equation}
\dot{C}_{10}+ \gamma _{10} C_{10}-i\sum_{j=1}^{N}\Omega _{Rj}^{\ast }C_{0j}e^{-i\Delta_j t}  =-\frac{i}{\hbar }\mathfrak{R}_{10},  
\label{C10 dotted}
\end{equation}%
\begin{equation}
\dot{C}_{0j} - \gamma _{0j}
C_{0j}-i\Omega _{Rj}C_{10}e^{i\Delta_j t} =-\frac{i}{\hbar }\mathfrak{R}_{0j},
\label{C0j dotted}
\end{equation}%
where the relaxation constants are related to the EM field and qubit
relaxation constants in the Lindbladian Eq.~(\ref{Lindbladian at 0 T}) by
\begin{equation}
\gamma _{00}=0,\ \gamma _{10}=\frac{\mu }{2},\ \gamma _{0j}=\frac{\gamma _{j}%
}{2}.  \label{relaxation constants}
\end{equation}%
The noise properties are given by%
\begin{equation}
\overline{\mathfrak{R}_{\alpha n}^{\ast }\left( t^{\prime }\right) \mathfrak{%
R}_{\beta m}\left( t^{\prime \prime }\right) }=\hbar ^{2}\delta _{\alpha
\beta }\delta _{nm}D_{\alpha n,\alpha n}\delta \left( t^{\prime }-t^{\prime
\prime }\right) ,  \label{noise properties}
\end{equation}%
\begin{equation}
D_{00,00}=\sum_{j=1}^{N}\gamma _{j}\overline{\left\vert C_{0j}\right\vert
^{2}}+\mu \overline{\left\vert C_{10}\right\vert ^{2}},\ D_{10,10}=0,\
D_{0j,0j}=0.  \label{D_abcd}
\end{equation}%

To include elastic relaxation (pure dephasing) in the Lindbladian Eq.~(\ref{Lindbladian at 0
T}) we need to add the term \cite{fain}%
\begin{equation*}
L^{\left( el\right) }(\hat{\rho})=-\Sigma _{j}[\frac{\gamma _{j}^{\left(
el\right) }}{2}(\hat{\sigma}_{zj}\hat{\sigma}_{zj}^{\dagger }\hat{\rho}+\hat{%
\rho}\hat{\sigma}_{zj}\hat{\sigma}_{zj}^{\dagger }-2\hat{\sigma}%
_{zj}^{\dagger }\hat{\rho}\hat{\sigma}_{zj})],
\end{equation*}%
where $\hat{\sigma}_{zj}=\hat{\sigma}_{zj}^{\dagger }=\left\vert
1_{j}\right\rangle \left\langle 1_{j}\right\vert -\left\vert
0_{j}\right\rangle \left\langle 0_{j}\right\vert $ and $\gamma _{j}^{\left(
el\right) }$ is an elastic relaxation constant. Recent analysis \cite{tokman2020,chen2021} shows that for two-level qubits the elastic processes
can be included by making the following replacements in the expressions for $%
\gamma _{0j}$ and $D_{0j,0j}$: $\gamma _{0j}\Longrightarrow \gamma _{0j}+$ $%
\gamma _{j}^{\left( el\right) }$, $D_{0j,0j}\Longrightarrow
D_{0j,0j}+2\gamma _{j}^{\left( el\right) }\overline{\left\vert
C_{0j}\right\vert ^{2}}$. These relationships lead to standard relaxation
timescales of populations $T_{1j}=\frac{1}{\gamma _{j}}$ and coherence $%
T_{2j}=\frac{1}{\frac{1}{2T_{1j}}+\gamma _{j}^{\left( el\right) }}$ \cite%
{fain}. Therefore, including pure dephasing processes leads to
corrections in the last of Eqs.~(\ref{relaxation constants}) and the last of
Eqs.~(\ref{D_abcd}), namely%
\begin{equation}
\gamma _{0j}=\frac{\gamma _{j}}{2}+\gamma _{j}^{\left( el\right) },\
D_{0j,0j}=2\gamma _{j}^{\left( el\right) }.\   \label{corrections due to edp}
\end{equation}

Taking into account Eqs.~(\ref{corrections due to edp}), it is easy
to show that for any set of elastic scattering rates Eqs.~(\ref{C00 dotted}%
)-(\ref{C0j dotted}) conserve the norm:
\begin{equation}
\sum_{j=1}^{N}\overline{\left\vert C_{0j}\right\vert ^{2}}+\overline{%
\left\vert C_{10}\right\vert ^{2}}+\overline{\left\vert C_{00}\right\vert
^{2}}=1;  \label{norm condition}
\end{equation}%
and Eqs.~(\ref{C10 dotted})-(\ref{C0j dotted}) preserve the following
relationship which includes only the rates of \textit{inelastic} relaxation:
\begin{equation}
\frac{d}{dt}\left( \sum_{j=1}^{N}\overline{\left\vert C_{0j}\right\vert ^{2}}%
+\overline{\left\vert C_{10}\right\vert ^{2}}\right) =-\sum_{j=1}^{N}\gamma
_{j}\overline{\left\vert C_{0j}\right\vert ^{2}}-\mu \overline{\left\vert
C_{10}\right\vert ^{2}}.  \label{relationship}
\end{equation}

If pure dephasing processes can be neglected and the reservoir temperature is much lower than the optical transition frequency (in energy units), we always have $D_{0j,0j} = D_{10,10} = 0$, which, together with $\overline{\mathfrak{R}_{10}} = \overline{\mathfrak{R}_{0j}} = 0$, allows one to neglect the contribution of noise sources $\mathfrak{R}_{10}$ and $\mathfrak{R}_{0j}$ when calculating observables; see \cite{tokman2020,chen2021}. In this case, Eqs.~(\ref{C00 dotted})-(\ref{C0j dotted}) can be considered an improved version of 
the Weisskopf-Wigner approximation, because they not only include dissipation as imaginary parts of eigenenergies but also  conserve
the norm of the state vector; see Eq.~(\ref{norm condition}). 
It is worth noting that within these approximations, such a simple method as a modified Weisskopf-Wigner approach has the same accuracy as the Lindblad formalism for the density matrix. 
See in this respect the paper \cite{mollow1975} which played an important part in formulating the SSE approach. \\


\section{Quantum dynamics and emission spectrum of an ensemble of qubits in a dissipative nanocavity}

\subsection{Analytic solution for quantum dynamics in a nonuniform cavity field}

Here we consider a typical low-Q plasmonic cavity with a field decay time much shorter than dissipation times in qubits $T_{(1,2)j}$. In this case the dissipation is dominated by the field decay, and we can put $\gamma_{0j} \approx 0$ in Eqs.~(\ref{C00 dotted})-(\ref{C0j dotted}). Furthermore, considering the low-temperature limit (as compared to the optical frequency) we can put $\gamma_{00} = D_{0j,0j} = D_{10,10} = 0$, which, together with $\overline{\mathfrak{R}_{10}} = \overline{\mathfrak{R}_{0j}} = 0$, allows one to neglect the effect of noise terms $\mathfrak{R}_{10}$ and $\mathfrak{R}_{0j}$ \cite{tokman2020,chen2021}. The resulting coupled equations for the probability amplitudes $C_{10}$ and $C_{0j}$ read 
\begin{equation}
\dot{C}_{10}+ \frac{\mu}{2} C_{10}-i \sum_{j=1}^{N}\Omega _{Rj}^{\ast }C_{0j}e^{-i\Delta_j t}  =0,  
\label{C10-2}
\end{equation} 
\begin{equation}
\dot{C}_{0j} - i\Omega _{Rj}C_{10}e^{i\Delta_j t} =0,
\label{C0j-2}
\end{equation}
whereas the solution for the amplitude of the ground state is 
\begin{equation}
C_{00}(t) = C_{00}(t = 0) -\frac{i}{\hbar } \int_0^t \mathfrak{R}_{00} dt,
 \label{C00-2}
\end{equation}
so that $\overline{C_{00}(t) - C_{00}(t = 0)} = 0$. The value of $\overline{C_{00}^2(t)}$ can be also determined directly from the conservation law  (\ref{norm condition}), but we will need Eq.~(\ref{C00-2}) when calculating the emission spectrum below. 

These equations can be immediately solved for an ensemble of qubits with the same transition frequencies but with different Rabi frequencies since they are located in a nonuniform field of a nanocavity. We will keep the assumption that the nanocavity field decays much faster than the qubit excitation. The case of different transition frequencies is considered in the next section. We can put $\Delta_j = 0$ in Eqs.~(\ref{C10-2}), (\ref{C0j-2}) and introduce the new variable 
\begin{equation}
F =  \sum_{j=1}^{N}\Omega _{Rj}^{\ast }C_{0j},  
\label{F}
\end{equation} 
which yields 
\begin{equation}
\dot{C}_{10}+ \frac{\mu}{2} C_{10}-i F =0,  
\label{C10-3}
\end{equation} 
\begin{equation}
\dot{F} - i\Omega _{N}^2 C_{10} =0,
\label{C0j-3}
\end{equation}
where
\begin{equation}
\Omega_N^2 = \sum_{j=1}^{N} |\Omega _{Rj}|^2
\label{omegaN}
\end{equation}
is a collective Rabi frequency. The initial conditions $C_{10}(0) = F(0) = 0$ give a trivial steady state solution, and there is an infinite number of states corresponding to $F = 0$. 

Seeking the solution $\propto e^{\Gamma t}$ gives 
\begin{equation}
\left(
\begin{array}{c}
C_{10} \\
F 
\end{array}
\right) = e^{-\frac{\mu}{4} t} \left[ A e^{i \Sigma t} \left(
\begin{array}{c} 
1 \\
K_1 
\end{array} \right) + B e^{-i \Sigma t} \left(
\begin{array}{c} 
1 \\
K_2 
\end{array} \right) \right], 
\label{sol-ident}
\end{equation}
where 
\begin{equation}
K_{1,2} = \pm \Sigma - i \frac{\mu}{4}, \; \Sigma = \sqrt{\Omega_N^2 - \frac{\mu^2}{16}}
\label{k1,2} 
\end{equation}
and the constants $A$ and $B$ are given by the initial conditions 
\begin{equation}
A = \frac{K_2 C_{10}(0) - F(0)}{K_2 - K_1}, \;  B = \frac{F(0) - K_1 C_{10}(0)}{K_2 - K_1}, 
\nonumber
\end{equation}
and
\begin{equation} 
F(0) = \sum_{j=1}^N \Omega_{Rj}^* C_{0j}(0).
\nonumber
\end{equation}

Similarly, from Eq.~(\ref{C0j-2}) when $\Delta_j = 0$ we obtain 
\begin{equation}
C_{0j}(t) = C_{0j}(0) + i \Omega_{Rj}  \int_0^{t} C_{10}(t') dt'.
 \label{C0j-3}
\end{equation}
Using the solution for $C_{10}$ which follows from Eq.~(\ref{sol-ident}),  
\begin{equation}
C_{10}(t) =  \left[ C_{10}(0) \left( \cos\Sigma t - \frac{\mu}{4 \Sigma} \sin\Sigma t \right) + i \frac{F(0)}{\Sigma} \sin\Sigma t \right]
 e^{-\frac{\mu}{4} t}, 
\label{c10-infty} 
\end{equation}
we arrive at
\begin{equation}
C_{0j}(\infty) = C_{0j}(0) - \Omega_{Rj} \frac{F(0)}{\Omega_N^2}. 
 \label{C0j-4}
\end{equation}
Note that Eq.~(\ref{C0j-4}) is valid for any $\mu$. 

At long times one always has $C_{10}(\infty) = F(\infty) = 0$. Therefore, for the initial state satisfying the condition
\begin{equation}
\frac{C_{0j}(0)}{C_{0i}(0)} = \frac{ \Omega_{Rj}}{\Omega_{Ri}}, 
 \label{init1}
\end{equation}
all energy stored initially in the qubit system is radiated away over a short cavity decay time $\sim 1/\mu$. Such a state is the generalization of the bright Dicke state (see, e.g., \cite{shammah2018,reitz2022}) to an ensemble of quantum emitters strongly coupled to a spatially nonuniform field of a plasmonic cavity.

Consider an arbitrary initial state:
\begin{equation}
\left\vert \Psi (0) \right\rangle = \left(
\begin{array}{c}
C_{00}(0) \\
C_{10}(0) \\
C_{01}(0) \\
... \\
C_{0j}(0) \\
... \\
C_{0N}(0) 
\end{array}
\right) = \left(
\begin{array}{c}
C_{00}(0) \\
0 \\
0 \\
... \\
0 \\
... \\
0 
\end{array}
\right) +  \left( 
\begin{array}{c}
0 \\
C_{10}(0) \\
C_{01}(0) \\
... \\
C_{0j}(0) \\
... \\
C_{0N}(0) 
\end{array}
\right). 
\label{arb-init}
\end{equation}
We are interested in the subset of equations for variables $C_{10}$ and $C_{0j}$ that end up being separated from the ground state. Since the system is linear, we can split the last column on the right-hand side of Eq.~(\ref{arb-init})  into two components: 
\begin{equation}
 \left(
\begin{array}{c}
C_{10}(0) \\
C_{01}(0) \\
... \\
C_{0j}(0) \\
... \\
C_{0N}(0) 
\end{array}
\right) = \left( 
\begin{array}{c}
0 \\
C_{01}(0) - \Omega_{R1} \frac{F(0)}{\Omega_N^2} \\
... \\
C_{0j}(0)  - \Omega_{Rj} \frac{F(0)}{\Omega_N^2}  \\
... \\
C_{0N}(0) - \Omega_{RN} \frac{F(0)}{\Omega_N^2} 
\end{array}
\right) + \left( 
\begin{array}{c}
C_{10}(0)  \\
 \Omega_{R1} \frac{F(0)}{\Omega_N^2} \\
... \\
 \Omega_{Rj} \frac{F(0)}{\Omega_N^2}  \\
... \\
 \Omega_{RN} \frac{F(0)}{\Omega_N^2} 
\end{array}
\right) . 
\label{arb-init2}
\end{equation}
It is easy to see that the first column on the right-hand side of Eq.~(\ref{arb-init2}) corresponds to a stationary (dark) state with $C_{10} = F = 0$. The second column gives rise to the bright state found before. As a result, we obtain
\begin{equation}
C_{10}(\infty) = 0, \; C_{0j}(\infty) = C_{0j}(0) - \Omega_{Rj} \frac{\sum_{m=1}^N \Omega_{Rm}^* C_{0m}(0) }{\sum_{j=1}^N| \Omega_{Rj}|^2}.
\label{arb-init3}
\end{equation}
Then from Eq.~(\ref{norm condition})  the amplitude of the ground state is given by 
\begin{equation}
|C_{00}(\infty)|^2 = 1 - \sum_{j = 1}^N \left|  C_{0j}(0) - \Omega_{Rj} \frac{\sum_{m=1}^N \Omega_{Rm}^* C_{0m}(0) }{\sum_{j=1}^N| \Omega_{Rj}|^2} \right|^2.
\label{arb-init4}
\end{equation}

It is clear from Eqs.~(\ref{arb-init3}) and (\ref{arb-init4}) that if the number $J$ of initially excited qubits is much smaller than the total number of qubits, $J \ll N$, then the change of the initial quantum state of the qubit ensemble is of the order of $J/N$. Therefore, an ensemble of ground-state qubits effectively shields an arbitrary initial state of a relatively small group of excited qubits from coupling to the cavity field. The shielding is due to formation of an entangled dark state in which the destructive interference leads to decoupling of the many-body state from the cavity field, even though each qubit remains strongly coupled to the quantum field of a cavity. 


\begin{figure}[htb]
\includegraphics[width=0.5\textwidth]{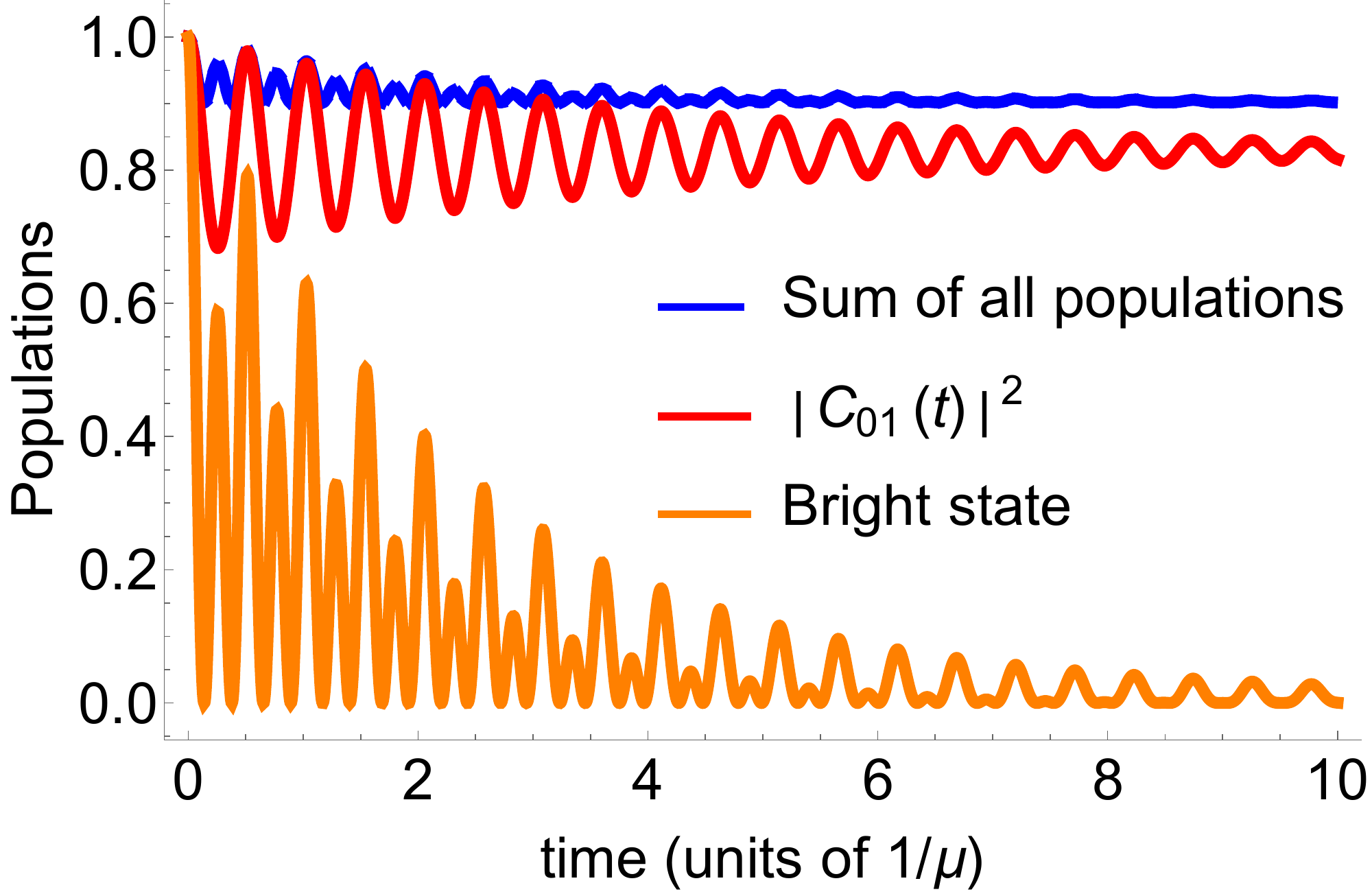}
\caption{Time evolution of the populations for an ensemble of $N = 21$ qubits in the nanocavity formed by the metallic sphere of radius $R = 10$ nm with its center located at $z_0 = 12$ nm above the substrate; see Appendix A for the field distribution. The molecules are assumed to be distributed uniformly within a circle of radius 10 nm on the substrate, with the center of the circle at the cavity axis $\rho = 0$. The effective cavity volume is 50 nm$^3$, the transition dipole moment is 10 Debye, the Rabi frequency at $\rho = 0$ is 120 meV, and the cavity decay time is $1/\mu = 20$ fs. Top blue curve: The sum of the occupation probabilities of all qubits $\sum_{j = 1}^N |C_{0j}|^2$ when only the qubit in the center of the cavity is excited, i.e., $C_{01}(0) = 1$; middle red curve: the occupation $|C_{01}(t)|^2$ of the initially excited qubit; bottom orange curve: the sum of all occupation probabilities when the qubits were initally prepared in the bright state (\ref{init1}).   }
\label{fig2main}
\end{figure}


The following numerical example in Fig.~\ref{fig2main} illustrates the formation of an entangled dark state in an ensemble of qubits in a nonuniform field of a nanocavity. To have an explicit analytic expression for the nanocavity field distribution we use the model described in Appendix A: a metallic sphere over a metallic substrate, where the metallic sphere can represent a nanoparticle or an apex of a nanotip as in recent strong-coupling experiments \cite{chikkaraddy2016, park2016, pelton2018, park2019}. Of course, for a realistic geometry of any specific experiment, one can calculate the field distribution numerically, using, e.g., a finite-element solver. For our example, we take the sphere of radius $R = 10$ nm with its center located at $z_0 = 12$ nm above the substrate. We will use the line charge approximation (\ref{lineApprox}) for the electric field of a cavity mode, which is an excellent approximation to the exact formula, as one can see from the middle plot in Fig.~\ref{zDependence}. Let us take $N = 21$ qubits distributed uniformly on the substrate at distances from $\rho = 0$ to 10 nm from the $z$-axis. It is obvious from Fig.~\ref{zDependence} that they experience very different cavity field amplitudes. First we consider an arbitrary initial state which is neither bright nor dark. Let us assume for definiteness that only one qubit located at the maximum field $\rho = 0$ is initially excited, i.e., its initial probability amplitude $C_{01}(t=0) = 1$, whereas all other qubits are in the ground state. The subsequent excitation of this qubit as described by $|C_{01}(t)|^2$ is shown as the middle red curve in Fig.~\ref{fig2main}, whereas the sum of populations of all qubits is the top blue curve. As is obvious from the picture, after the bright state component of the initial state is radiated away over a short time of a few $1/\mu$, the system remains in an entangled dark state which is decoupled from the cavity mode and has a lifetime determined by relaxation constants of the qubits. This can also be verified by calculating $F(t)$ from Eq.~(\ref{F}) which approaches zero over the same timescale. Only a few per cent ($\sim 1/N$) of the total excitation energy is radiated away.  This result remains qualitatively the same when we vary the distribution of the initial excitation; only the fraction of the radiated energy changes. 

The dynamics changes if the system was initially prepared {\it exactly} in the bright state described by Eq.~(\ref{init1}). In this case {\it all}  initial excitation is radiated away over the time of the order of a few $1/\mu$. The bottom (orange) curve in Fig.~\ref{fig2main} shows the behavior of the sum of all qubit populations when the system starts from the bright state. 

\subsection{The emission spectrum}

Detecting the radiation from quantum emitters placed in nanocavities is one of
the most straightforward ways to study their quantum dynamics \cite%
{Scully1997, madsen2013, lodahl2015, chikkaraddy2016, pelton2018, park2019}.
The power spectrum received by the detector can be calculated as \cite%
{madsen2013,Scully1997}%
\begin{equation*}
P\left( \nu \right) =A\cdot S\left( \nu \right)
\end{equation*}%
where%
\begin{equation}
S\left( \nu \right) =\frac{1}{\pi }\mathrm{Re}\int_{0}^{\infty }d\tau
e^{i\nu \tau }\int_{0}^{\infty }dtK\left( t,\tau \right) ,  \label{S(v)}
\end{equation}%
\begin{equation}
K=\left\langle \Psi \left( 0\right) \right\vert \hat{c}^{\dagger }\left(
t\right) \hat{c}\left( t+\tau \right) \left\vert \Psi \left( 0\right)
\right\rangle ;  \label{K}
\end{equation}%
$\hat{c}^{\dagger }\left( t\right) $ and $\hat{c}\left( t\right) $ are
Heisenberg creation and annihilation operators for the cavity field, $\Psi
\left( 0\right) $ is an initial state of the system. The coefficient $A$ is
determined by the cavity design, spatial structure of the cavity field, and
detector properties.

These equations indicate that to calculate the power spectrum one has to
solve the Heisenberg-Langevin equations for the operators $\hat{c}\left(
t\right) $ and $\hat{c}^{\dagger }\left( t\right) $ \cite{Scully1997} and
evaluate the correlator including averaging over the noise statistics, $%
K\Rightarrow $ $\overline{\left\langle \Psi \left( 0\right) \right\vert \hat{%
c}^{\dagger }\left( t\right) \hat{c}\left( t+\tau \right) \left\vert \Psi
\left( 0\right) \right\rangle }$. However, the Heisenberg-Langevin equations
are nonlinear in the strong-coupling Rabi oscillations regime for a
single-photon field. Therefore, it is more convenient to utilize the
solution of the linear stochastic equation~(\ref{S eq}) for the state vector. The corresponding procedure is described in \cite{parametric} where we prove that the correlator $K\left( t,\tau \right)$ can be calculated as 
\begin{equation}
K\left( t,\tau \right) =\overline{\left\langle \ \Phi \left( t,\tau \right)
\right. \left\vert \Psi _{C}\left( t+\tau \right) \right\rangle }.
\label{K bar}
\end{equation}%
Here   $\Psi _{C}\left( t+\tau \right) = \hat{c}\Psi
\left( t+\tau \right) $, where $\Psi \left( t+\tau \right) $ is the solution
to the stochastic Schr\"{o}dinger equation (\ref{S eq}) on the time interval $\left[ 0,t+\tau \right]
$ with initial condition $\left\vert \Psi \left( 0\right) \right\rangle $; $\Phi \left( t,\tau \right) $ is the solution to Eq.~(\ref{S eq})  on the time interval $%
\left[ t,t+\tau \right] $ with initial condition $\Psi _{C}\left( t\right) $, and $\Psi _{C}\left( t\right) = \hat{c}\Psi
\left( t \right)$, where $\Psi
\left( t \right)$ is also the solution Eq.~(\ref{S eq}) but over the time interval $\left[ 0,t\right] $. The overbar in Eq.~(\ref{K bar}) denotes averaging over the statistics of noise sources, which according to  the Langevin
approach is equivalent to averaging over the reservoir degrees of freedom  \cite{Landau1965}.

Now we apply this formalism to calculate the emission spectrum of an excited qubit in an ensemble of ground-state qubits. Since we just want to illustrate how the formation of an entangled dark state suppresses the emission from the cavity, we can simplify algebra and consider identical Rabi frequencies: $\Omega_{Rj} = \Omega_R$, $\Omega_N^2 = N \Omega_R^2$. If needed, a more cumbersome analytic solution for the spectrum can also be readily obtained for an arbitrary distribution of Rabi frequencies using the state vector derived in the previous subsection.

 As before, we will
solve for the evolution over the intermediate timescales when only the field
dissipation has to be taken into account. Consider an initial state in which
only one qubit is excited, $\left\vert \Psi (0) \right\rangle = \left\vert 0 \right\rangle \left\vert 1_1 \right\rangle \Pi _{m=2}^{N}\left\vert
0_{m}\right\rangle$.

As
usual, we seek the solution of the stochastic equation for the state vector
in the form of Eq.~(\ref{Psi}). From Eqs.~(\ref{Psi}) and (\ref{S eq}) one can get%
\begin{equation}
\Psi _{C}\left( t\right) =\hat{c}\Psi \left( t\right) =C_{10}\left( t\right)
\left\vert 0\right\rangle \Pi _{j=1}^{N}\left\vert 0_{j}\right\rangle
\label{psiC}
\end{equation}

According to the above procedure, we need to find the solution of Eqs.~(\ref{C10-2})-(\ref{C00-2}) with initial condition (\ref{psiC}) at the time
interval $[t,t+\tau]$. One can see that Eqs.~(\ref{C10-2}),(\ref{C0j-2}) have a trivial zero solution, whereas Eq.~(\ref{C00-2}) yields
\begin{equation}
\ \Phi \left( t,\tau \right) =\left[  C_{10}\left( t\right) -\frac{i}{\hbar }e^{-i\frac{\omega }{2}\tau
}\int_{0}^{\tau } \mathfrak{R}_{00}\left(
t+t^{\prime }\right) dt^{\prime }\right] \left\vert 0\right\rangle \Pi
_{j=1}^{N}\left\vert 0_{j}\right\rangle.  \label{Phi}
\end{equation}%
Substituting Eqs.~(\ref{psiC}) and (\ref{Phi}) into Eq.~(\ref{K bar}) and
taking into account that the term linear with respect to the noise source
gives zero upon averaging, we obtain%
\begin{equation}
K\left( t,\tau \right) = C_{10}^{\ast }\left(
t\right) C_{10}\left( t+\tau \right).  \label{k ave}
\end{equation}%
Using Eq.~(\ref{sol-ident}) for the function $C_{10}\left( t\right)$ we get 
\begin{equation}
K\left( t,\tau \right) =\frac{\left\vert \Omega _{R}\right\vert^{2}}{\Sigma
^{2}}e^{-\frac{\mu }{4} \tau }e^{-\frac{\mu }{2}%
t}\sin \left( \Sigma t\right) \sin \left[ \Sigma \left( t+\tau \right) \right] .
\label{k(tao,t)}
\end{equation}%
The resulting power spectrum in Eq.~(\ref{S(v)}) is given by%
\begin{equation*}
S\left( \nu \right) =\frac{8\left\vert \Omega _{R}\right\vert ^{2}}{\pi
\mu \left( \mu ^{2}+16\Sigma ^{2}\right) }\mathrm{Re}\frac{\mu -2i \nu
}{\left( \frac{\mu }{4}-i\nu  \right)^{2}+\Sigma ^{2}}. 
\end{equation*}
Taking into account the fact that we solved the problem in the interaction picture, the measured spectrum is obtained by replacing $\nu \Rightarrow  \nu - \omega$. Using also Eq.~(\ref{k1,2}), we obtain
\begin{equation}
S\left( \nu \right) =\frac{1}{2\pi }\frac{\left\vert \Omega
_{R}\right\vert ^{2}}{\left( \left( \nu -\omega \right) ^{2}-\left(
N\left\vert \Omega _{R}\right\vert ^{2}-\frac{\mu ^{2}}{8}\right) \right)
^{2}+\frac{\mu ^{2}}{4}\left( N\left\vert \Omega _{R}\right\vert ^{2}-\frac{%
\mu ^{2}}{16}\right) } .  \label{power spectrum}
\end{equation}%
Under the condition $\mu \ll 2\left\vert \Omega _{R}\right\vert \sqrt{N}$ the spectrum is simplified: 
\begin{equation*}
S( \nu) =\frac{1}{2\pi }\frac{\left\vert \Omega _{R}\right\vert ^{2}}{\left(
\left( \nu -\omega \right) ^{2}-N\left\vert \Omega _{R}\right\vert
^{2}\right) ^{2}+\frac{\mu ^{2}}{4}N\left\vert \Omega _{R}\right\vert ^{2}},
\end{equation*}%
i.e., the spectrum consists of two well-resolved lines shifted with respect
to $\omega $ by $\pm \left\vert \Omega _{R}\right\vert \sqrt{N}$, with the
maximum value $S_{\max }\left( \pm \left\vert \Omega _{R}\right\vert \sqrt{N}%
\right) =\frac{1}{\pi }\frac{2}{N\mu ^{2}}$ and linewidth $\sim \frac{\mu }{2%
}$. The dependence $S_{\max }\propto \frac{1}{N}$ reflects the destructive interference 
effect described above: the probability of the photon emission by a
qubit scales as $P_{rad}\approx \frac{1}{N}$.


\begin{figure}[htb]
\includegraphics[width=0.5\textwidth]{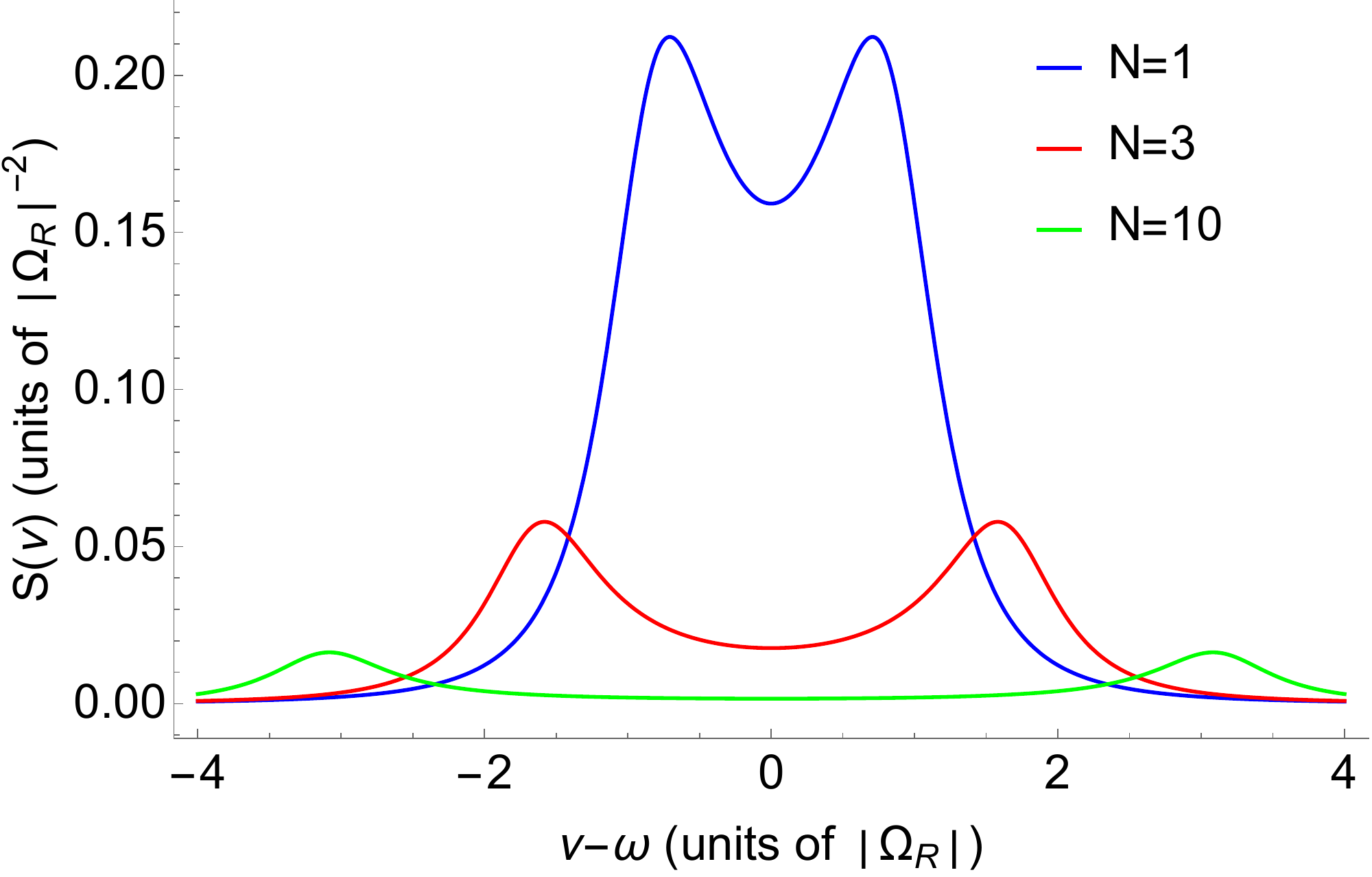}
\caption{Normalized emission spectra given by Eq.~(\protect\ref{power
spectrum}) for three values of $N$ and the cavity decay rate $\protect\mu/2
= \Omega_R$. The height of the peaks scales as $1/N$. }
\label{spectra}
\end{figure}

This behavior is illustrated in Fig.~\ref{spectra} which shows the emission
spectra given by Eq.~(\ref{power spectrum}) for three different qubit
numbers $N$ and the cavity decay rate $\mu/2 = \Omega_R$. The most interesting result here is not the splitting of the spectrum which is an obvious consequence of strong coupling, but the fact that the peak intensity (the height of the peaks) gets suppressed with increasing $N$ as $1/N$.  

As we already pointed out, the dissipation-driven transition of a system into a dark state is not surprising by itself and has been studied before for various systems; see, e.g., the formation of subradiant states in the Dicke superradiance problem \cite{gegg2018} or quantum dots in a plasmonic cavity \cite{gray2015,gray2016}.  It is nontrivial, however, that in our case of a strongly coupled $N$-qubit system, the amount of energy loss from the system before it goes into the dark state approaches zero as $1/N$ due to destructive interference from unexcited qubits. It is also convenient that we have a complete analytic solution describing the effect. 


\section{Many-qubit systems with different transition frequencies}

In this section we consider an ensemble of qubits with a large spread of
transition frequencies interacting with a spatially nonuniform cavity mode. This is usually the case for quantum dots where the inhomogeneous broadening is related to the dispersion of the dot's sizes. We will assume that the
inhomogeneous broadening dominates: 
\begin{equation}
\frac{\mu}{4 \Delta_m} \ll 1, 
\label{dominates}
\end{equation}
where $\Delta_m$ is the half-width of the inhomogeneous broadening. We will show below that under strong-coupling conditions the inhomogeneous broadening leads to long-period pulsations of individual qubit populations but does not prevent the formation of a collective dark state decoupled from the cavity mode, as long as the {\it collective Rabi frequency} $\Omega_N$ in Eq.~(\ref{omegaN}) remains larger than $\Delta_m$. 

It follows from Eq.~(\ref{C0j-2}) that 
\begin{equation}
C_{0j} = C_{0j}(0) + i \Omega_{Rj}  \int_0^{t} C_{10}(\tau) e^{i\Delta_j \tau} d\tau,
 \label{C0j-5}
\end{equation}
which can be substituted into Eq.~(\ref{C10-2}) to obtain
\begin{equation}
\dot{C}_{10} + \frac{\mu}{2} C_{10} = i \sum_{j=1}^N \Omega_{Rj}^* C_{0j}(0) e^{-i\Delta_j t}  - \int_0^t  \sum_{j=1}^N |\Omega_{Rj}|^2  C_{10}(\tau) e^{i\Delta_j (\tau - t)} d\tau.
 \label{C10-5}
\end{equation}

Now we introduce the Laplace transform,
\begin{equation}
C_p = \int_0^{\infty} C_{10}(t) e^{-pt} dt, \; C_{10}(t) = \frac{1}{2 \pi i}  \int_{x-i\infty}^{x+i\infty} C_{p} e^{pt} dp.
\nonumber
\end{equation}
Since the functions $\sum_{j=1}^N \Omega_{Rj}^* C_{0j}(0) e^{-i\Delta_j t}$  and   $\sum_{j=1}^N |\Omega_{Rj}|^2 e^{-i\Delta_j t}$ do not grow as $t \rightarrow \infty$, we can assume Re$[p] > 0$ and therefore $x > 0$.
Laplace transforming Eq.~(\ref{C10-5}) gives
\begin{equation}
pC_p - C_{10}(0) + \frac{\mu}{2} C_p = i F_p -  C_p D_p,
 \label{C10-8}
\end{equation}
where 
\begin{equation}
F_p =  \int_{0}^{\infty} \left( \sum_{j=1}^N \Omega_{Rj}^* C_{0j}(0) e^{-(i\Delta_j +p) t} \right) dt  = \sum_{j=1}^N \frac{ \Omega_{Rj}^* C_{0j}(0) }{i\Delta_j +p}, 
\nonumber
\end{equation}
\begin{equation}
D_p=  \int_{0}^{\infty} \sum_{j=1}^N |\Omega_{Rj}|^2 e^{-(i\Delta_j + p) t} dt  = \sum_{j=1}^N \frac{ |\Omega_{Rj}|^2 }{i\Delta_j +p}.  
\nonumber
\end{equation}
Solving Eq.~(\ref{C10-8}) gives
\begin{equation}
C_{10}(t)  = \frac{1}{2 \pi i}  \int_{x-i\infty}^{x+i\infty} \frac{C_{10}(0) + \sum_{j=1}^N \frac{ \Omega_{Rj}^* C_{0j}(0) }{i\Delta_j +p}}{p + \frac{\mu}{2} + \sum_{j=1}^N \frac{ |\Omega_{Rj}|^2 }{i\Delta_j +p}} e^{pt} dp.  
 \label{C10-9}
\end{equation}
The functions $C_{0j}(t)$ are determined by substituting Eq.~(\ref{C10-9}) into Eq.~(\ref{C0j-5}). 

The behavior of the function $C_{10}(t)$ is determined by zeros of the denominator of the integrand in Eq.~(\ref{C10-9}):
\begin{equation}
C_{10}(t) \rightarrow \sum_k A_k e^{p_{0k} t},
\nonumber
\end{equation}
 where $p_{0k}$ are the solutions of equation 
 \begin{equation}
\left( p + \frac{\mu}{2}\right) \Pi_j^N (i\Delta_j + p)  + \sum_{j=1}^N |\Omega_{Rj}|^2 \Pi_{k \neq j}^N (i\Delta_k + p) = 0. 
\label{poles} 
\end{equation}
Equation (\ref{poles}) determines a set of $N+1$ normal modes for the system of Eqs.~(\ref{C10-2}) and (\ref{C0j-2}) after the replacement $C_{0j}(t) e^{-i\Delta_j t} \rightarrow C_{0j}(t)$ which eliminates explicit time dependence. The Laplace transform is especially convenient in the limit of a continuous spectrum, see Appendix B. The dynamics of the populations of individual qubits should include the beatnotes with characteristic periods $T \sim \frac{\pi N}{\Delta_m}$. At the same time, as long as the collective Rabi frequency $\Omega_N$ remains greater than the inhomogeneous linewidth, strong coupling still leads to the formation of a collective dark state in which only a small fraction $\sim 1/N$ of the initial excitation energy is radiated away whereas the sum of all qubit populations remains approximately constant and close to its initial value.


\begin{figure}[htb]
\includegraphics[width=0.5\textwidth]{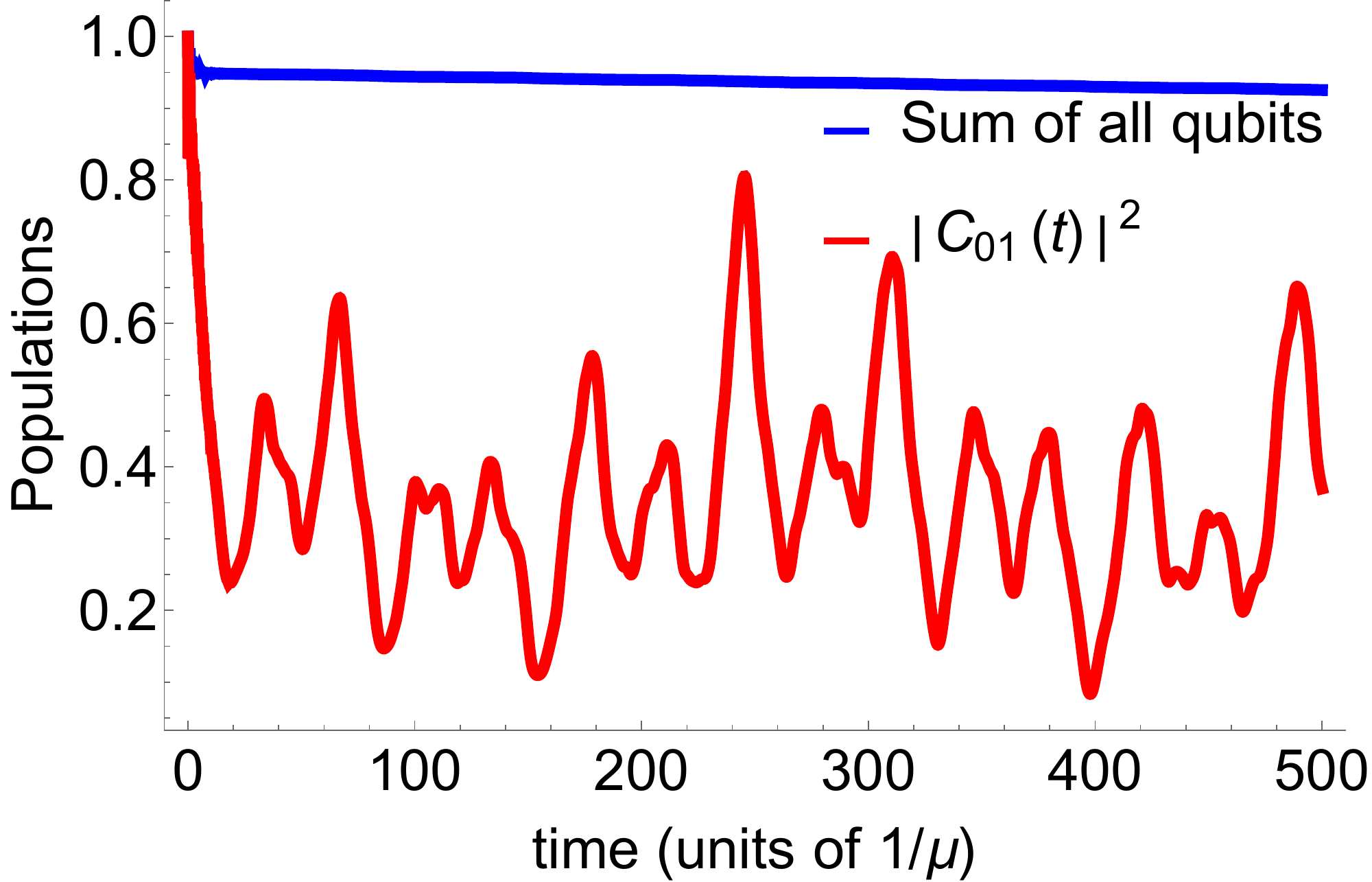}
\caption{Time evolution of the populations for an ensemble of $N = 41$ qubits with transition frequencies distributed pseudo-randomly in the range $\pm \Delta_m = 50$ meV around resonance with a cavity mode. The cavity decay, Rabi frequency distribution, geometry, and spatial distribution are the same as for the example in Fig.~\ref{fig2main}. Top blue curve: the sum of the occupation probabilities of all qubits $\sum_{j = 1}^N |C_{0j}|^2$ when only one qubit in the center of the cavity is excited initially, i.e., $C_{01}(0) = 1$; bottom red curve: the occupation $|C_{01}(t)|^2$ of the initially excited qubit.    }
\label{fig3main}
\end{figure}



\begin{figure}[htb]
\includegraphics[width=0.6\textwidth]{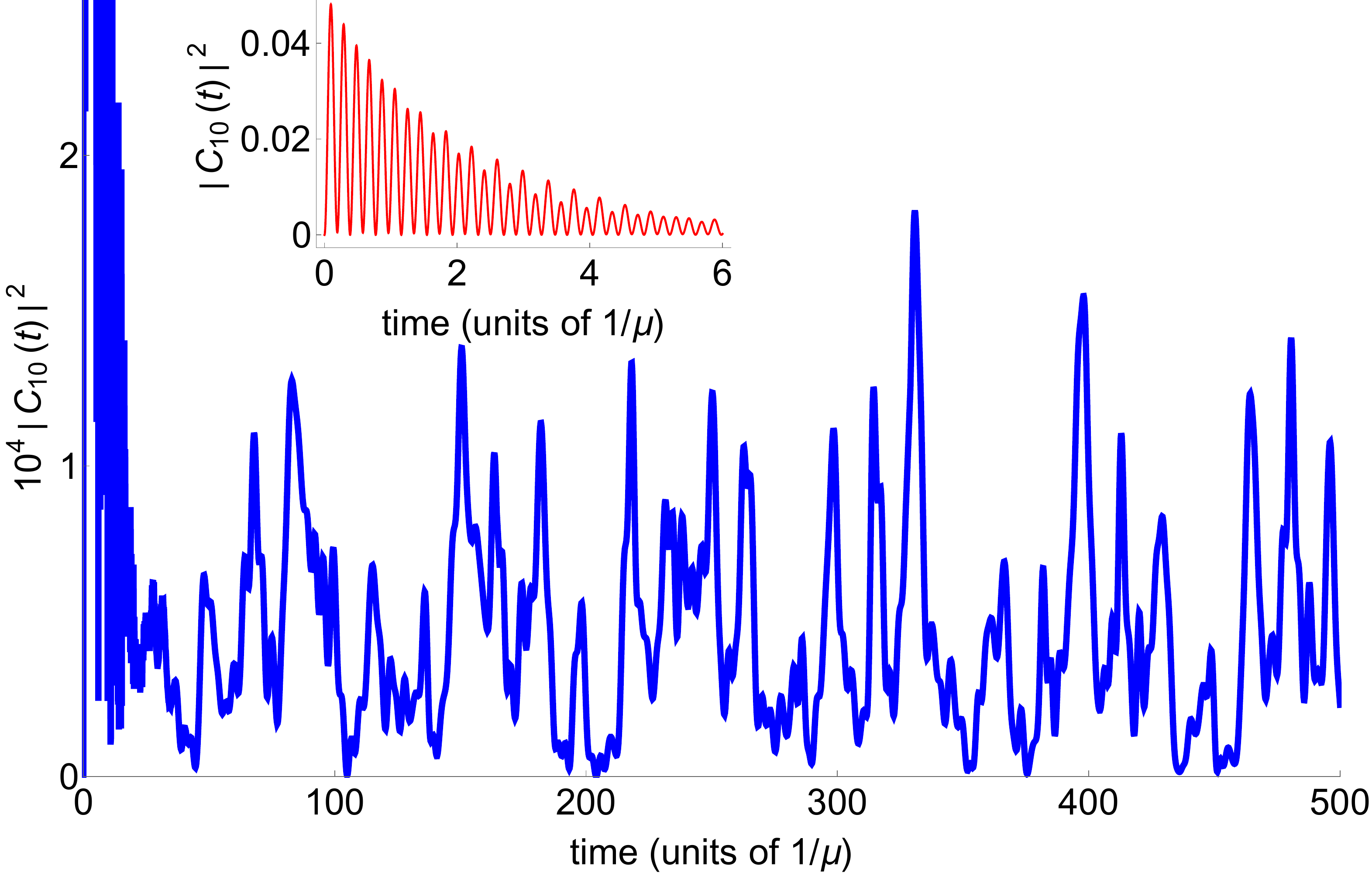}
\caption{Excitation probability of the cavity mode, $|C_{10}(t)|^2$, for the same conditions as in Fig.~\ref{fig3main}. Inset: same for a short initial time interval, showing initial relaxation of the cavity field and Rabi oscillations.   }
\label{fig4main}
\end{figure}


We illustrate this dynamics by solving numerically the set of Eqs.~(\ref{C10-2}) and (\ref{C0j-2}) for particular values of the parameters. One example is shown  in  Figs.~\ref{fig3main} and \ref{fig4main}. Here we consider $N = 41$ qubits with transition frequencies distributed pseudo-randomly in the range $\pm \Delta_m = 50$ meV around resonance with a cavity mode, which corresponds to typical spread of frequencies of semiconductor quantum dots. The geometry and spatial distribution are the same as for the example in Fig.~\ref{fig2main}. The cavity decay time is again 20 fs, i.e., $\mu = 33$ meV and the Rabi frequency in the center of the cavity is 120 meV. As is clear from the figures, over a very short initial time of the order of several $1/\mu$ a small $\sim 1/N$ fraction of the initial excitation energy is radiated away and the entangled dark state is established. After that, individual qubit populations undergo slow quasi-chaotic oscillations, as expected from a system of coupled oscillators with incommensurate frequencies, whereas the sum of all populations remains almost constant except for a very slow decay with characteristic timescale of $ > 10^4$ $1/\mu$. This decay is due to a small residual coupling to a cavity mode: as one can see from  the long-time dynamics in Fig.~\ref{fig4main}, the cavity mode maintains quasi-chaotic oscillations at a very low level of $\sim 10^{-4}$. Eventually, the relaxation of individual qubits which we neglected here will  kick in, typically over ps timescales at room temperatures and ns to $\mu$s scale at low temperatures.


\begin{figure}[htb]
\includegraphics[width=0.5\textwidth]{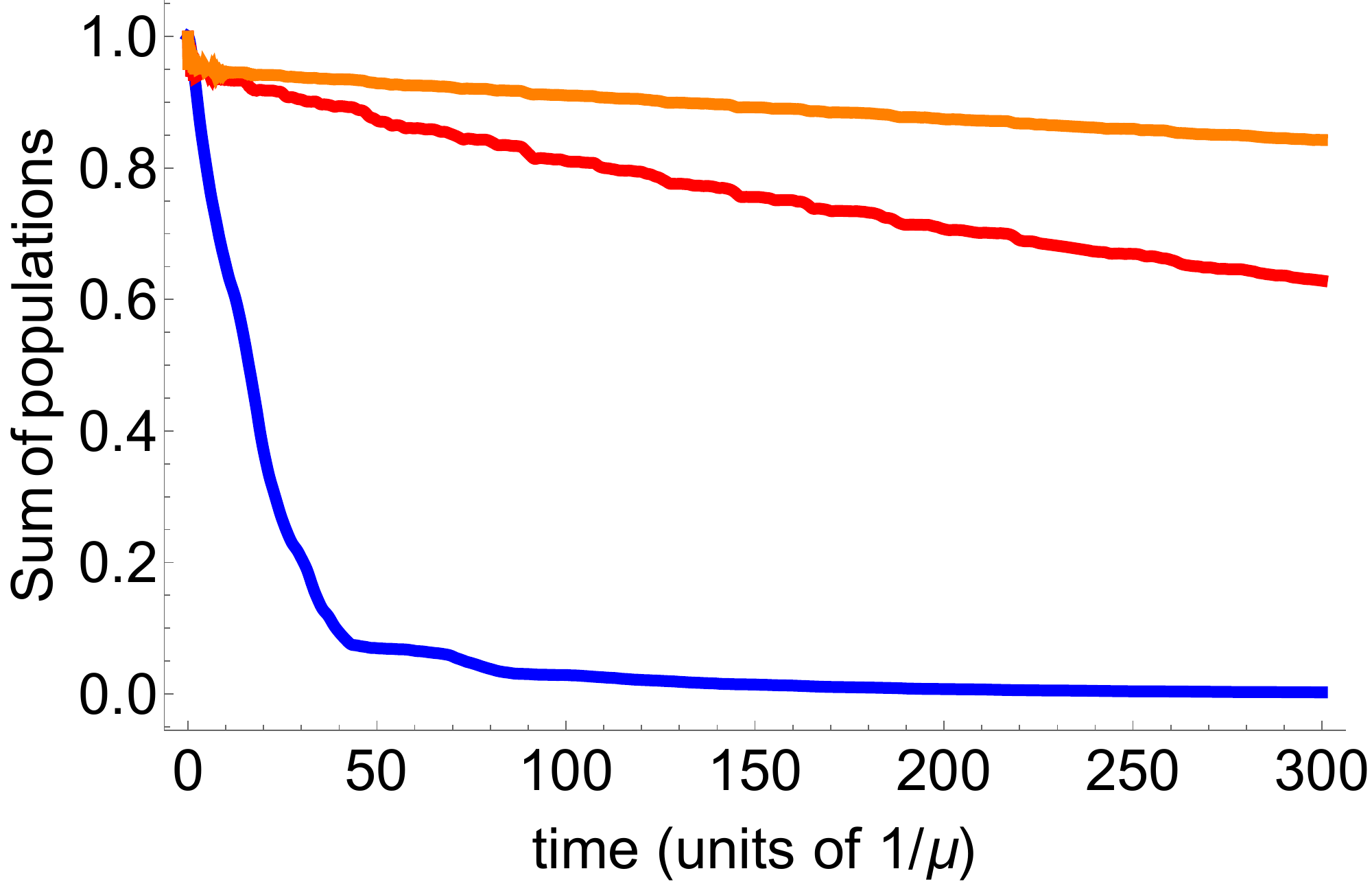}
\caption{Time evolution of the sum of the occupation probabilities of $N = 41$ qubits $\sum_{j = 1}^N |C_{0j}|^2$ with transition frequencies distributed pseudo-randomly in the range $\pm \Delta_m = 90$ meV around resonance with a cavity mode. The cavity decay time, Rabi frequency distribution, geometry, and spatial distribution are the same as for the example in Fig.~\ref{fig2main}. Only one qubit in the center of the cavity is initially excited, i.e., $C_{01}(0) = 1$. Three curves correspond to three different collective Rabi frequencies $\Omega_N$. Top orange curve: $\Omega_N = 540$ meV, middle red curve: $\Omega_N = 270$ meV, bottom blue curve: $\Omega_N = 27$ meV.   }
\label{fig-variedRabi}
\end{figure}

The initial stage of relaxation to the dark state ($\mu t \leq 20$) is modulated by fast Rabi oscillations that are not even visible in Fig.~\ref{fig3main} but can be seen in the inset of Fig.~\ref{fig4main}. The subsequent slow beatnote oscillations of individual qubit populations vary from qubit to qubit and between different random realizations of the distribution of transition frequencies, but the qualitative picture remains the same. The beatnote oscillations become strictly periodic when the transition frequencies are separated by the same frequency interval, but this would be an unrealistic situation. 

If the collective Rabi frequency $\Omega_N$ becomes smaller than $\Delta_m$, the decay of the sum of the occupation probabilities of all qubits $\sum_{j = 1}^N |C_{0j}|^2$ accelerates. This is illustrated in Fig.~\ref{fig-variedRabi} which shows the evolution of the sum of populations for three values of the Rabi frequency at the cavity center: $\Omega_R(0) = 120$ meV (top curve), 60 meV (middle curve), and 6 meV (bottom curve, which correspond the the values of the collective Rabi frequency $\Omega_N = 540$ meV, 270 meV, and 27 meV, respectively. There is an obvious shortening of the decay time when $\Omega_N$ becomes much smaller than the total spread of transition frequencies determined by $2 \Delta_m = 180$ meV.

With increasing spectral density of qubits the periods of beatnotes increase, eventually leading to a continuous spectrum of inhomogeneous broadening, where further analytic insights can be obtained, especially for the photon mode dynamics which is not significantly affected by beatnote oscillations. Some limiting cases are described in Appendix B.


\subsection{Electron states in quantum wells}

Here we consider a multilevel quantum-confined
electron system such as electron states in a quantum well or perhaps in a quantum wire or a multilevel quantum dot.  In this case optical
transitions occur generally between two groups of electron energy states,
for example between electron states in the conduction band and valence band.
Let's take zero energy between these two groups and denote positive energies
in the conduction band as $W_{j}$ (Latin indices) and negative energies in
the valence band as $-W_{\alpha }$ (Greek indices). The frequencies of the
optical transitions are%
\begin{equation}
\omega _{j\alpha }=\frac{W_{j}+W_{\alpha }}{\hbar }.  \label{fot}
\end{equation}%
We won't consider here the intraband optical transitions within each group,
e.g. $\alpha \iff \beta $ or $m\iff n$, although the formalism below can be
easily extended to include them.

The RWA Hamiltonian is
\begin{equation}
\hat{H}=\hbar \omega \left( \hat{c}^{\dagger }\hat{c}+\frac{1}{2}\right)
+\sum_{j=1}^{J}W_{j}\hat{a}_{j}^{\dagger }\hat{a}_{j}-\sum_{\alpha
=1}^{A}W_{\alpha }\hat{a}_{\alpha }^{\dagger }\hat{a}_{\alpha }-\hbar
\sum_{j=1}^{J}\sum_{\alpha =1}^{A}\left[ \left( \Omega _{R;j\alpha }\hat{a} 
_{j}^{\dagger }\hat{a}_{\alpha }\hat{c}+\Omega _{R;j\alpha }^*\hat{a}_{\alpha
}^{\dagger }\hat{a}_{j}\hat{c}^{\dagger }\right) \right] 
\label{RWA hamiltonian}
\end{equation}%
where $\Omega _{R;j\alpha }=\frac{\mathbf{d}_{\alpha j}\mathbf{\cdot E}}{%
\hbar }.$

It is again convenient to work in the interaction picture where 
\begin{equation}
\hat{H} = -\hbar
\sum_{j=1}^{J} \sum_{\alpha =1}^{A} \Omega _{R;j\alpha }\hat{a}_{j}^{\dagger } \hat{a}_{\alpha }\hat{c} e^{i \Delta_{j\alpha} t} + {\rm h. c.}  
\label{RWA-2}
\end{equation}
where $\Delta_{j\alpha} = \omega_{j\alpha} - \omega$. Note that the electric-dipole-forbidden transitions are eliminated by values $d_{j\alpha} = 0$.

Instead of the excitation and deexcitation operators for a qubit that are
specific to a two-level system, $\hat{\sigma}^{\dagger }$and $\hat{\sigma}$,
it is easier to introduce standard creation and annihilation operators of
the fermion states. Therefore, the states that were denoted as $\left\vert
0_{j\alpha }\right\rangle $ and $\left\vert 1_{j\alpha }\right\rangle $ when
using the operators $\hat{\sigma}^{\dagger }$ and $\hat{\sigma}$ become $%
\left\vert 0_{j}\right\rangle \left\vert 1_{\alpha }\right\rangle $ and $%
\left\vert 1_{j}\right\rangle \left\vert 0_{\alpha }\right\rangle $ when
using standard fermion operators.

We consider again lowest-energy states corresponding to zero- or
single-photon excitations:
\begin{eqnarray}
\Psi & = & C_{00}\left\vert 0\right\rangle \Pi _{j=1}^{J}\left\vert
0_{j}\right\rangle \Pi _{\alpha =1}^{A}\left\vert 1_{\alpha }\right\rangle
+C_{10}\left\vert 1\right\rangle \Pi _{j=1}^{J}\left\vert 0_{j}\right\rangle
\Pi _{\alpha =1}^{A}\left\vert 1_{\alpha }\right\rangle  \notag \\
& + & \sum_{j,\alpha }^{N,A}C_{0j\alpha }\left\vert 0\right\rangle
\left\vert 1_{j}\right\rangle \left\vert 0_{\alpha }\right\rangle \Pi
_{m\neq j}^{J}\left\vert 0_{m}\right\rangle \Pi _{\beta \neq \alpha
}^{A}\left\vert 1_{\beta }\right\rangle.  \label{spe}
\end{eqnarray}%

Equations for the probability amplitudes $C_{10}$ and $C_{0j\alpha}$ within the stochastic Schr\"{o}dinger equation formalism become 
\begin{equation}
\dot{C}_{10} + \frac{\mu}{2} C_{10} - i \sum_{j=1}^J  \sum_{\alpha =1}^{A} \Omega _{R;j\alpha }^* C_{0j\alpha} e^{-i \Delta_{j\alpha} t} = 0,
 \label{C10-qw}
\end{equation}
\begin{equation}
\dot{C}_{0j\alpha} -i  \Omega _{R;j\alpha } C_{10} e^{i \Delta_{j\alpha} t} = 0.
 \label{C0j-qw}
\end{equation}
If spin states are degenerate, pairs $\{j,\alpha\}$ corresponding to different spin states $\{j_{\downarrow},\alpha_{\downarrow} \}$ and $\{j_{\uparrow},\alpha_{\uparrow} \}$ have to be taken into account separately in Eqs.~(\ref{C10-qw}) and (\ref{C0j-qw}).

To proceed, we assign the number $s = 1,..., J\times A$ to each pair $\{j,\alpha\}$ and therefore reduce the problem to the one already solved in this section. We leave the straightforward algebra to the reader. 
The most interesting result, in our opinion, is still the formation of a long-lived entangled dark state decoupled from the cavity field when the collective Rabi frequency $\left( \sum_{j=1}^J  \sum_{\alpha =1}^{A} |\Omega _{R;j\alpha }|^2 \right)^{1/2}$ exceeds the width of the inhomogeneous broadening $|\Delta_{j\alpha}|_{max}$. 


\section{Nonclassical multiphoton states in dissipative strongly-coupled systems} 

Many of the results obtained in previous sections for single-photon excitations, in particular the formation of dark entangled qubit states decoupled from the cavity field, can be generalized to arbitrary multiphoton excitations which correspond to $N \geq M$ and $M > 1$ in Eqs.~(\ref{a1}) and (\ref{a5}). To avoid cumbersome algebra, consider an example of equal Rabi frequencies and exact resonance, when one can put $\Omega_{Rj} = \Omega_R$ and $\Delta_j = 0$ in the Hamiltonian (\ref{hint}). This is not a critical assumption and it can be avoided at the expense of more complicated final expressions. Within the stochastic  equation for the state vector, any group of probability amplitudes with a fixed value of $M = n + p$ is described by the following system of equations,
\begin{equation}
\left( \frac{d}{dt} + \gamma _{np\alpha_p} \right) C_{np\alpha_p} - i \left( \Omega_R \sqrt{n+1} \sum_{\alpha_{p-1}}^{p} C_{(n+1)(p-1)\alpha_{p-1}} + \Omega_R^* \sqrt{n} \sum_{\alpha_{p+1}}^{N-p} C_{(n-1)(p+1)\alpha_{p+1}} \right)
= \mathfrak{R}_{np\alpha_p}(t),
 \label{b1}
\end{equation}
where
\begin{equation}
\overline{\mathfrak{R}_{np\alpha_p}(t)  \mathfrak{R}_{n'p'\alpha_p'}^*(t') } = \hbar ^{2} \delta( t- t') D_{np\alpha_p; n'p'\alpha_p'}. 
\label{b2}
\end{equation}
The lower index in the sums shows the type of a subset and the upper index shows the number of elements in the sum. Equation (\ref{b1}) implies that the subsets $\alpha_{p-1}$ and $\alpha_{p+1}$ are related to subset $\alpha_{p}$ through
\begin{equation}
\label{b3} 
|p, \alpha_p \rangle =  \hat{\sigma}_{j_{p-1}}^{\dagger} \left\vert p-1, \alpha_{p-1} \right\rangle, \;  |p, \alpha_p \rangle =  \hat{\sigma}_{j_{p+1}} \left\vert p+1, \alpha_{p+1} \right\rangle
\end{equation}
where each pair $\alpha_p, \alpha_{p-1}$ or $\alpha_p, \alpha_{p+1}$ corresponds to a certain value of the qubit index: $j_{p-1}$ or $j_{p+1}$.  Each subset $\alpha_p$ corresponds to a certain finite number of subsets  $\alpha_{p-1}$ or $\alpha_{p+1}$ which contribute to the summation in Eq.~(\ref{b1}). 

In the general case the presence of noise source terms $\mathfrak{R}_{np\alpha_p}$ couples the groups with different values of $M$. However, in the strong coupling regime such a noise-induced coupling scales as a small parameter $\frac{\gamma _{np\alpha_p} }{\Omega_R} \ll 1$ and therefore can be included perturbatively. A similar perturbative approach has been developed for nonlinear strong coupling of electron-photon-phonon systems \cite{parametric}. 

Furthermore, for high enough photon frequencies $\hbar \omega \gg T$, one can assume zero temperature of dissipative reservoirs. At optical frequencies this is true even at room temperature. In this case the method of determining relaxation rates $\gamma _{np\alpha_p}$ and correlators $D_{np\alpha_p; n'p'\alpha_p'}$ is described in Sec.~III. Assuming in addition that field dissipation is dominant in a nanocavity, we obtain
\begin{equation}
\gamma _{np\alpha_p} = n \frac{\mu}{2},
\label{b4}
\end{equation}
\begin{equation}
D_{n  p\alpha_p; n'p'\alpha_p'} = \langle n | \langle p, \alpha_p | \delta \hat{L}(\hat{\rho})_{\hat{\rho} = \overline{\left\vert \Psi \right\rangle
\left\langle \Psi \right\vert }} | p', \alpha_p' \rangle | n' \rangle = \mu \delta_{pp'} \delta_{\alpha_p \alpha_p'} \sqrt{(n+1)(n'+1)} \times \overline{ C_{(n+1) p\alpha_p} C_{(n'+1) p'\alpha_p'}^* },
\label{b5}
\end{equation}
where the operator $\delta \hat{L}(\hat{\rho})$ is determined by the last term in Eq.~(\ref{delta LO}). It follows from Eq.~(\ref{b5}) that nonzero autocorrelators of noise terms inside the group with a fixed value of $M = n + p$ are determined by averages of the amplitudes $\overline{ C_{(n+1) p\alpha_p} C_{(n+1) p\alpha_p}^* }$ from the group with $M \Rightarrow M+1$:
$$
D_{n  p\alpha_p; np\alpha_p} = \mu (n+1) \overline{ C_{(n+1) p\alpha_p} C_{(n+1) p\alpha_p}^* };
$$
whereas, nonzero cross-correlators coupling the groups with different $M = n+p$ and $M' = n'+p'$ are determined by the amplitudes $\overline{ C_{(n+1) p\alpha_p} C_{(n'+1) p\alpha_p}^* }$ from the groups with $M \Rightarrow M+1$ and $M' \Rightarrow M'+1$:
$$
D_{n  p\alpha_p; n'p\alpha_p} = \mu \sqrt{ (n+1)(n'+1)} \overline{ C_{(n+1) p\alpha_p} C_{(n'+1) p\alpha_p}^* }.
$$

Therefore, for low-temperature reservoirs the coupling between blocks with different $M$ exists only in the downward direction. The maximum value of $M$ is determined by the initial energy of the system; thermal excitations above initial $M$ are impossible. Within the group corresponding to maximum $M$ all correlators $\overline{\mathfrak{R}_{np\alpha_p}(t)  \mathfrak{R}_{n'p'\alpha_p'}^*(t') }$ are equal to zero and therefore one can neglect the noise terms in Eq.~(\ref{b1}) for this group as they don't affect the observables. The noise terms in lower-$M$ groups affect how the deexcitation proceeds across all possible relaxation channels (as, e.g., in \cite{parametric}). At the same time the relaxation rate of the states in the highest-$M$ group is determined only by the values of $\gamma _{np\alpha_p} = n \frac{\mu}{2}$. 

These properties allow us to obtain intuitive analytic results describing quantum dissipative multiqubit dynamics at low reservoir temperature. For example, consider the states in the  highest-$M$ group where we can put $\mathfrak{R}_{np\alpha_p} = 0$ in Eq.~(\ref{b1}) and take into account Eq.~(\ref{b4}). This gives 
\begin{equation}
\begin{array}{c} 
\left( \frac{d}{dt} +  n \frac{\mu}{2} \right) C_{n(M-n)\alpha_{M-n}} - i \left( \Omega_R \sqrt{n+1} \sum_{\alpha_{M-n-1}}^{M-n} C_{(n+1)(M-n-1)\alpha_{M-n-1}} \right. \\ 
\left. 
+ \Omega_R^* \sqrt{n} \sum_{\alpha_{M-n+1}}^{N-M+n} C_{(n-1)(M-n+1)\alpha_{M-n+1}} \right)
= 0,
\end{array}
 \label{b6}
\end{equation}
where $n = 0,1, ... , M$. 

The main difficulty with solving Eqs.~(\ref{b6}) is related to the rules imposed by Eq.~(\ref{b3}),which dictate how each element of the subset 
$ \alpha_{M-n}$ is related to the elements of subsets $ \alpha_{M-n \mp 1}$ which enter the sums $\sum_{\alpha_{M-n-1}}^{M-n}( ...)$ and $\sum_{\alpha_{M-n+1}}^{N-M+n} (...)$, respectively. However, one avoids this complication when finding complex energy eigenvalues by summing each of Eqs.~(\ref{b6}) over all subsets $ \alpha_{M-n}$. This results in the following equations for the variables 
$$
F_n =     \sum_{\alpha_{M-n}}^{ \mathcal{C}_N^{M-n}} C_{n(M-n) \alpha_{M-n}}:
$$
\begin{equation}
\left( \frac{d}{dt} +  n \frac{\mu}{2} \right) F_n - i \left( \Omega_R \sqrt{n+1} (N-M+n+1)F_{n+1} + \Omega_R^* \sqrt{n} (M-n+1) F_{n-1} \right) = 0.
 \label{b7}
\end{equation}

For example, consider the case of $M = 2$. Seeking $F_n \propto e^{\Gamma t}$ we obtain
\begin{equation}
\begin{pmatrix}
\Gamma & -i(N-1) \Omega_R & 0 \\
-i 2 \Omega_R^* & \Gamma + \frac{\mu}{2} & -i \sqrt{2} \Omega_R \\ 
0 & -i \sqrt{2} \Omega_R^* & \Gamma + \mu 
\end{pmatrix} 
 \left( 
\begin{array}{c}
F_0 \\
F_1 \\
F_2 
\end{array}
\right) = 0,
\label{b8}
\end{equation}
which gives 
\begin{equation}
\Gamma \left( \Gamma + \frac{\mu}{2} \right)  ( \Gamma + \mu) + 2 N |\Omega_R|^2 \Gamma + 2 (N-1) |\Omega_R|^2 \mu = 0. 
 \label{b9}
\end{equation}

When $N \gg 1$, equation (\ref{b9}) can be factorized:
\begin{equation}
( \Gamma + \mu) \left[ \Gamma \left( \Gamma + \frac{\mu}{2} \right)   + 2 N |\Omega_R|^2 \right] = 0, 
 \nonumber
\end{equation}
which gives 
\begin{equation}
\Gamma_{1,2} \approx - \frac{\mu}{4} \pm i  \left(  2 N |\Omega_R|^2 \Gamma - \frac{\mu^2}{16} \right)^{1/2} , \; \Gamma_3 \approx - \mu.  
 \label{b10}
\end{equation}
It is easy to see that the roots $\Gamma_{1,2} $ describe evolution of coupled 1-photon and 0-photon states, 
$$
\Psi_{n = 0,1} = \Psi_{n = 0} + \Psi_{n = 1} =     \sum_{\alpha_{2}}^{ \mathcal{C}_N^{2}} C_{02 \alpha_{2}} | 0 \rangle |2, \alpha_2 \rangle +    \sum_{\alpha_{1}}^{ \mathcal{C}_N^{1}} C_{11 \alpha_{1}} | 1 \rangle |1, \alpha_1 \rangle,
$$
whereas root $\Gamma_3$ describes evolution of the 2-photon state,
$$
\Psi_{n = 2} =  C_{20 \alpha_{0}} | 2 \rangle  |0, \alpha_0 \rangle, \; {\rm where} \;  |0, \alpha_0 \rangle \equiv  |0_{qub} \rangle.
$$

Therefore, for a large number of qubits the 2-photon state evolves independently of other states and decays with decay rate $\mu$. At the same time, 1-photon and 0-photon states get entangled while oscillating with collective Rabi frequency $\approx  \left(  2 N |\Omega_R|^2 \Gamma - \frac{\mu^2}{16} \right)^{1/2}$ and decay with decay rate $\frac{\mu}{4}$. 

As the next example, we consider an initial state in which $M$ qubits are excited whereas the cavity field is in the vacuum state, i.e., 
$\Psi^{(0)} =  \sum_{\alpha_{M}}^{ \mathcal{C}_N^{M}} C_{0M \alpha_{M}}^{(0)} |0 \rangle |M,  \alpha_M \rangle$. The superscript $(0)$ denotes initial moment of time $t = 0$. An arbitrary initial state is a superposition of bight and dark initial states. Let's consider their evolution separately. 


\subsection{ Dark states} 

These are uncoupled from the cavity field and therefore are relatively long-lived, especially in the nanocavity QED context where the relaxation is dominated by the cavity field decay.  The dark states must satisfy the conditions
\begin{equation}
\sum_{\alpha_{M}}^{ N-M+1} C_{0M \alpha_{M}}^{(0)} = 0. 
\label{b11}
\end{equation}
Every element of the subset $\alpha_{M-1}$ in Eqs.~(\ref{b11}) is related to the elements of subset $\alpha_{M}$ in the sum $\sum_{\alpha_{M}}^{ N-M+1} (...)$ according to the rules of Eqs.~(\ref{b3}). It is easy to see that an initial state vector which satisfies the conditions $C_{(n>0)(M-n) \alpha_{M-n}}^{(0)} = 0$ and Eqs.~(\ref{b11}) remains constant with time, i.e., is a stationary solution of Eqs.~(\ref{b6}). 

Equations (\ref{b11}) contain $\mathcal{C}_N^{M-1}$ equations for $\mathcal{C}_N^{M}$ variables, i.e., the dark state conditions can be satisfied when $\mathcal{C}_N^{M} > \mathcal{C}_N^{M-1}$. This gives the condition for the existence of dark states: not more than half of the qubits can be initially excited, 
\begin{equation}
N \geq 2 M. 
\label{b12}
\end{equation}

The structure of a dark state can be visualized for a simple example, when $M = 2$ and $N = 4$. In this case the initial state vector is given by 
\begin{equation}
\begin{array}{c} 
\Psi^{(0)} = |0\rangle \left( C_{12}^{(0)} | 1 \rangle  | 1 \rangle  | 0 \rangle | 0 \rangle + C_{13}^{(0)} | 1 \rangle  | 0 \rangle  | 1 \rangle | 0 \rangle + C_{14}^{(0)} | 1 \rangle  | 0 \rangle  | 0 \rangle | 1 \rangle \right. \\
\left. + C_{23}^{(0)} | 0 \rangle  | 1 \rangle  | 1 \rangle | 0 \rangle + C_{24}^{(0)} | 0 \rangle  | 1 \rangle  | 0 \rangle | 1 \rangle + C_{34}^{(0)} | 0 \rangle  | 0 \rangle  | 1 \rangle | 1 \rangle \right), 
\end{array} 
\label{b13}
\end{equation}
where the ket before the parentheses is the photon state. Equations (\ref{b11}) become 
$$
C_{12}^{(0)}  + C_{13}^{(0)} + C_{14}^{(0)} = 0, \; C_{12}^{(0)}  + C_{23}^{(0)} + C_{24}^{(0)} = 0, \; C_{13}^{(0)}  + C_{23}^{(0)} + C_{34}^{(0)} = 0, \; C_{14}^{(0)}  + C_{24}^{(0)} + C_{34}^{(0)} = 0,
$$
which gives the dark state as 
\begin{equation}
C_{12}^{(0)} = C_{34}^{(0)}  = A, \;  C_{13}^{(0)} = C_{24}^{(0)} = B, \;  C_{14}^{(0)} =  C_{23}^{(0)} = C,  
\label{b14}
\end{equation}
and 
\begin{equation}
A + B + C = 0.   
\label{b15}
\end{equation}

Note that the dark states at any moment of time correspond to the trivial solution of Eqs.~(\ref{b7}): $F_n = 0$ for any $n$. Therefore, they cannot be analyzed with Eqs.~(\ref{b7}). 


\subsection{Bright states}

Obviously, one of the bright states is a completely symmetric state:
\begin{equation}
C_{0M \alpha_{M}}^{(0)} = {\rm const} = \frac{1}{\sqrt{\mathcal{C}_N^{M}}}. 
\label{b16}
\end{equation}
In this case due to symmetry we have $C_{n(M-n) \alpha_{M-n}} = \frac{F_n}{\mathcal{C}_N^{M-n}}$ at any moment of time. Such states are typical for the systems possessing permutational symmetry \cite{shammah2018}. Then from Eqs.~(\ref{b7}) we obtain that there is only one stationary state $F_n = 0$ for any $n$, which means that the energy of the state satisfying Eq.~(\ref{b16}) will be radiated away completely. 

The state given by Eq.~(\ref{b16})  is not the only bright state. Consider again the case of $M = 2$ and $N = 4$ for illustration. In this case the state vector at an arbitrary moment of time has the structure
\begin{eqnarray}
\displaystyle \Psi &=&  |0\rangle \left( C_{12} | 1 \rangle  | 1 \rangle  | 0 \rangle | 0 \rangle + C_{13} | 1 \rangle  | 0 \rangle  | 1 \rangle | 0 \rangle + C_{14} | 1 \rangle  | 0 \rangle  | 0 \rangle | 1 \rangle  + C_{23} | 0 \rangle  | 1 \rangle  | 1 \rangle | 0 \rangle + C_{24} | 0 \rangle  | 1 \rangle  | 0 \rangle | 1 \rangle + C_{34} | 0 \rangle  | 0 \rangle  | 1 \rangle | 1 \rangle \right) \nonumber  \\
&+& | 1 \rangle \left(  C_{1} | 1 \rangle  | 0 \rangle  | 0 \rangle | 0 \rangle + C_{2} | 0 \rangle  | 1 \rangle  | 0 \rangle | 0 \rangle + C_{3} | 0 \rangle  | 0 \rangle  | 1 \rangle | 0 \rangle + C_{4} | 0 \rangle  | 0 \rangle  | 0 \rangle | 1 \rangle \right)  + |2 \rangle C_{0} | 0 \rangle  | 0 \rangle  | 0 \rangle | 0 \rangle.  
\label{m2n4}  
\end{eqnarray}

Consider the following initial state: $ C_{14}^{(0)}  = -  C_{23}^{(0)} \neq 0$, $ C_{ij \neq 14,23}^{(0)} = 0$, $ C_{1,2,3,4}^{(0)} = 0$,  $C_{0}^{(0)} = 0$.  One can show that in this case at any moment of time $ C_{14}  = -  C_{23}$, $ C_{ij \neq 14,23}^{(0)} = 0$, $ C_1 = C_4 = - C_2 = -C_3$, $C_0 = 0$.  As a result, Eqs.~(\ref{b6}) yield the following equations, 
$$ 
\frac{d}{dt} C_{14} - 2 i \Omega_R C_1 = 0, \; \left( \frac{d}{dt} + \frac{\mu}{2} \right) C_1 - i \Omega_R^* C_{14} = 0,
$$
which describe decaying Rabi oscillations at frequency $\approx  \left(  2 |\Omega_R|^2 - \frac{\mu^2}{16} \right)^{1/2}$ resulting in a complete radiative energy loss with amplitude decay rate $\frac{\mu}{4}$. Formally, these expressions for the decay rate and Rabi frequency obtained using Eqs.~(\ref{b6}) are similar to those obtained from Eqs.~(\ref{b7}). However, it is easy to see that the above solution corresponds to the trivial solution of Eqs.~(\ref{b7}), i.e., $F_n = 0$ for all $n$, and therefore it cannot be derived from Eqs.~(\ref{b7}). 

Since the system is linear, an antisymmetric initial state of a more general form, 
$$ 
C_{12}^{(0)}  = -  C_{34}^{(0)}, \; C_{13}^{(0)} = - C_{24}^{(0)}, \;   C_{14}^{(0)} = - C_{23}^{(0)},
$$
is also bright. 

It is easy to see that any initial state of the type Eq.~(\ref{b13}) can always be split into two bright states (symmetric and antisymmetric one) and one dark state. For example, suppose that we initially excited one pair of qubits with probability of 1, i.e., $\Psi^{(0)} = | 0 \rangle  | 1 \rangle  | 1 \rangle | 0 \rangle | 0 \rangle$, where as always the first ket describes the photon state. This state can be represented as a sum of a symmetric bright state, 
\begin{equation}
\Psi_{bright}^{(s)}  = \frac{1}{6}| 0 \rangle \left(  | 1 \rangle  | 1 \rangle  | 0 \rangle | 0 \rangle + | 1 \rangle  | 0 \rangle  | 1 \rangle | 0 \rangle +  | 1 \rangle  | 0 \rangle  | 0 \rangle | 1 \rangle  + | 0 \rangle  | 1 \rangle  | 1 \rangle | 0 \rangle +  | 0 \rangle  | 1 \rangle  | 0 \rangle | 1 \rangle +  | 0 \rangle  | 0 \rangle  | 1 \rangle | 1 \rangle \right),
\nonumber 
\end{equation}
an asymmetric bright state, 
$$ 
\Psi_{bright}^{(as)}  = \frac{1}{2}| 0 \rangle \left(  | 1 \rangle  | 1 \rangle  | 0 \rangle | 0 \rangle -  | 0 \rangle  | 0 \rangle  | 1 \rangle | 1 \rangle \right), 
$$
and a dark state, 
\begin{equation}
\Psi_{dark}   = \frac{1}{6}| 0 \rangle \left( 2 | 1 \rangle  | 1 \rangle  | 0 \rangle | 0 \rangle - | 1 \rangle  | 0 \rangle  | 1 \rangle | 0 \rangle - | 1 \rangle  | 0 \rangle  | 0 \rangle | 1 \rangle - | 0 \rangle  | 1 \rangle  | 1 \rangle | 0 \rangle -  | 0 \rangle  | 1 \rangle  | 0 \rangle | 1 \rangle +  2 | 0 \rangle  | 0 \rangle  | 1 \rangle | 1 \rangle \right). 
\nonumber 
\end{equation}
One can see that $1/3$ of the original excitation energy goes to the dark state and is preserved until the qubit decay kicks in. The fraction of the preserved excitation increases if the initial state is closer to the dark state. For example, an initial state $\Psi^{(0)}  = \frac{1}{\sqrt{2}}| 0 \rangle \left(  | 1 \rangle  | 1 \rangle  | 0 \rangle | 0 \rangle +   | 0 \rangle  | 0 \rangle  | 1 \rangle | 1 \rangle \right)$ is a sum of a symmetric bright state, 
\begin{equation}
\Psi_{bright}^{(as)}  = \frac{1}{2\sqrt{2}}| 0 \rangle \left(  | 1 \rangle  | 1 \rangle  | 0 \rangle | 0 \rangle + | 1 \rangle  | 0 \rangle  | 1 \rangle | 0 \rangle +  | 1 \rangle  | 0 \rangle  | 0 \rangle | 1 \rangle + | 0 \rangle  | 1 \rangle  | 1 \rangle | 0 \rangle +  | 0 \rangle  | 1 \rangle  | 0 \rangle | 1 \rangle +  | 0 \rangle  | 0 \rangle  | 1 \rangle | 1 \rangle \right),
\nonumber 
\end{equation}
and a dark state, 
\begin{equation}
\Psi_{dark}   = \frac{1}{2\sqrt{2}} | 0 \rangle \left( | 1 \rangle  | 1 \rangle  | 0 \rangle | 0 \rangle - | 1 \rangle  | 0 \rangle  | 1 \rangle | 0 \rangle - | 1 \rangle  | 0 \rangle  | 0 \rangle | 1 \rangle  - | 0 \rangle  | 1 \rangle  | 1 \rangle | 0 \rangle -  | 0 \rangle  | 1 \rangle  | 0 \rangle | 1 \rangle +  | 0 \rangle  | 0 \rangle  | 1 \rangle | 1 \rangle \right). 
\nonumber 
\end{equation}
In this case $1/2$ of the original excitation energy goes into the dark state. 


\begin{figure}[htb]
\includegraphics[width=1.0\textwidth]{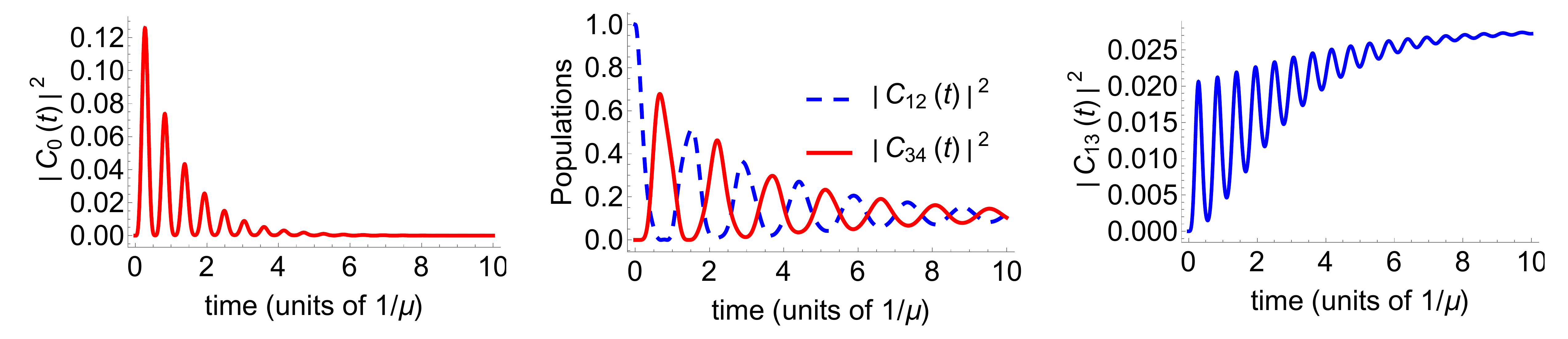}
\caption{An example of time evolution of populations for the $M = 2$, $N = 4$ state (\ref{m2n4}), when the two qubits are excited initially, namely $C_{12} = 1$ and all other coefficients are zero. The Rabi frequency $\Omega_R$ is 100 meV and cavity decay time $1/\mu = 20$ fs. Top panel: occupation probability of the two-photon state $|C_{0}(t)|^2$; middle panel: same for $|C_{12}(t)|^2$ and $|C_{34}(t)|^2$; bottom panel: same for $|C_{13}(t)|^2$. The dynamics of other  $|C_{ij}(t)|^2$ probabilities looks similar to that of $|C_{13}(t)|^2$.    }
\label{fig-m2n4}
\end{figure}



\begin{figure}[htb]
\includegraphics[width=0.45\textwidth]{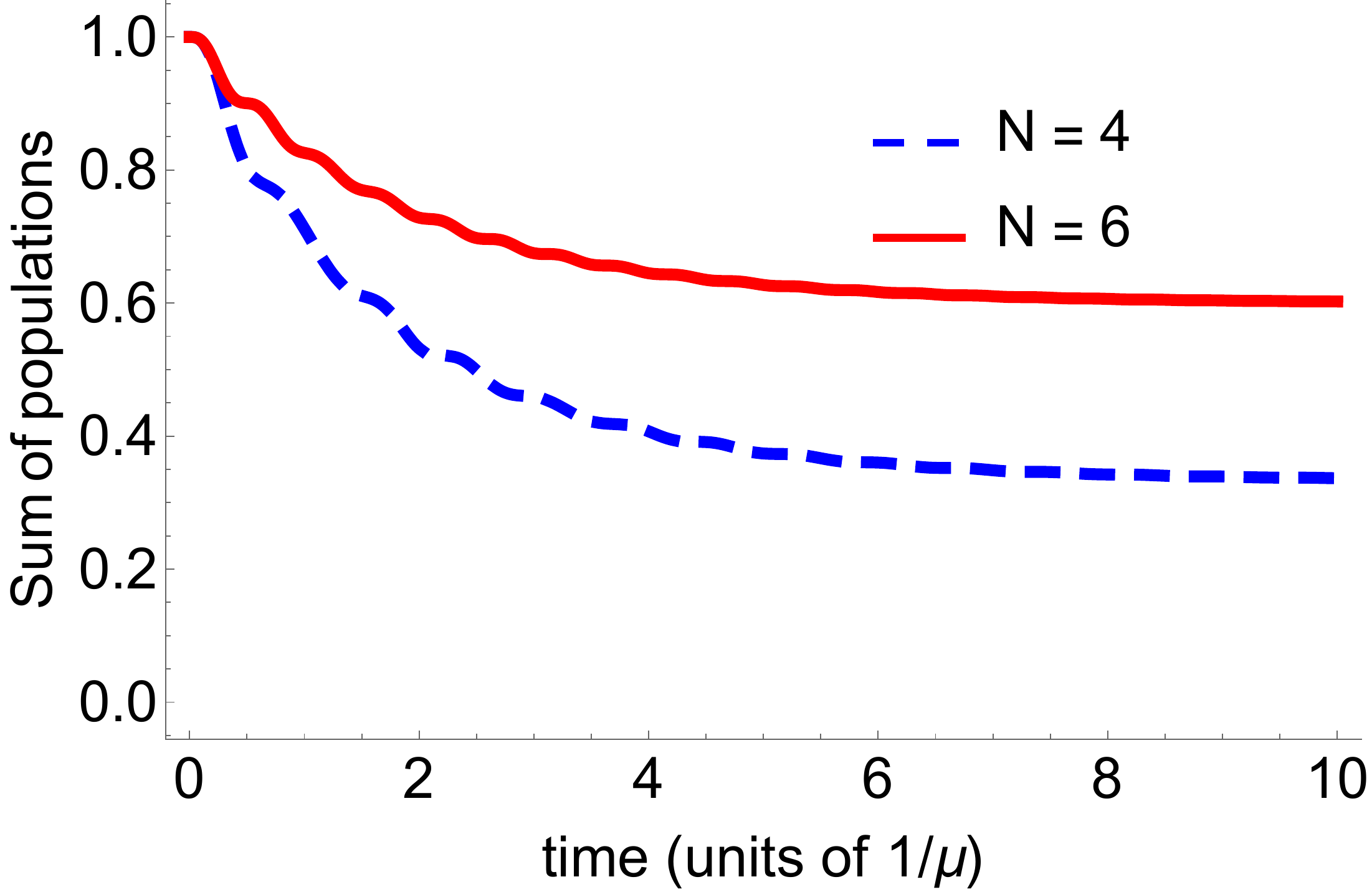}
\caption{Sum of all occupation probabilities for the $M = 2$, $N = 4$ state (\ref{m2n4}) (dashed blue curve) and the $M = 2$, $N = 6$ state (solid red curve) as a function of time, for the same initial conditions: two qubits are excited, namely $C_{12} = 1$ and all other coefficients are zero.    }
\label{fig-sum}
\end{figure}


Figures \ref{fig-m2n4} and \ref{fig-sum} illustrate this dynamics with a numerical example by solving Eqs.~(\ref{b6}) with the rules imposed by Eq.~(\ref{b3}) for the initial state  $\Psi^{(0)} = | 0 \rangle  | 1 \rangle  | 1 \rangle | 0 \rangle | 0 \rangle$ in which two qubits are excited with unit probability and all other coefficients are zero. This initial state is a mix of bright and dark states. As is clear from Fig.~\ref{fig-m2n4} plotted for the $M = 2$, $N = 4$ state given by Eq.~(\ref{m2n4}), the bright state part is radiated away over the time of several $1/\mu$, after which all occupations containing one or two photons, namely $|C_{j}(t)|^2$ where $j = 0,1,2,3,4$, approach zero whereas all two-qubit coefficients approach an entangled dark state decoupled from the cavity mode, in which the sum of all qubit populations is equal to 1/3 as predicted by our analytic theory; see the dashed blue curve in Fig.~\ref{fig-sum}. 

With increasing total number of qubits $N$ the fraction of the initial excitation which goes into the dark state increases rapidly, as illustrated with the $M = 2$, $N = 6$ example in Fig.~\ref{fig-sum}; see the solid red curve. This behavior is qualitatively similar to the case of single-photon excitations solved in the main text. 

If the experiment has a complete control over qubit excitations, one can switch between dark and bright states as needed; however, even in the case of no control the fact that a large or even dominant fraction of the initial excitation goes into a long-lived dark state makes low-Q plasmonic nanocavities more appealing for applications.

For large values of $m$ and $N$ the procedure of expanding an initial state into bright and dark states is unlikely to be simpler than direct solution of ordinary differential equations (\ref{b6}) obtained within the SSE method. However, there is a class of initial states for which this procedure is still the simplest. Consider the subset of states which don't have any common qubit and denote it as $|M, \tilde{\alpha}_M \rangle$. There are obviously $L = \frac{N}{M}$ of such states and we consider only the excitations where $L$ is integer. If only such states are excited initially and all initial amplitudes are the same and equal to $\frac{1}{\sqrt{L}}$, such states keep almost all their initial energy, especially for large $N-M \gg 1$: the amplitudes of states in $|M, \tilde{\alpha}_M \rangle$ approach $\frac{1}{\sqrt{L}} \frac{N-M}{N-M+1}$ whereas the amplitudes of all other states $|M, \alpha_M \rangle$ are excited from zero to the level of  $\frac{1}{\sqrt{L}} \frac{1}{N-M+1}$.


\section{Conclusions}

We found analytic solutions for the quantum dynamics of many-qubit systems strongly coupled to a quantized electromagnetic cavity mode, in the presence of decoherence and dissipation for both fermions and cavity photons.  Analytic or semi-analytic solutions are derived for a broad class of open quantum systems including identical qubits, an ensemble of qubits in a nonuniform nanocavity field with a broad distribution of coulping strengths and transition frequencies, and multi-level electron systems. The formalism is based on the stochastic equation  of evolution for the state vector, within Markov approximation for the relaxation processes and rotating wave approximation with respect to the optical transition frequencies. Although the stochastic Schr\"{o}dinger equation is typically used for numerical Monte-Carlo simulations, our version of this approach turned out to be convenient for the analytic theory.

We demonstrated in the analytic derivation that the interaction of an ensemble of qubits with a single-mode spatially nonuniform quantum field leads to 
 entangled states of practical importance, with destructive or constructive interference between the qubits depending on the initial excitation. In particular, if one or a small fraction of qubits were excited initially whereas the field was in the vacuum state, the subsequent relaxation drives the whole ensemble of qubits into an entangled dark state which is completely decoupled from the leaky cavity mode, even though each qubit remains strongly coupled to the field. It is nontrivial that only a small fraction $~ 1/N$ of the initial excitation energy is lost before the system goes into the dark state, where $N$ is the number of qubits in the ground state.

We found the conditions in which strong coupling overcomes the spread of transition frequencies of an ensemble of qubits or a multi-electron system and leads to formation  of a decoupled many-qubit dark state with conserved total excitation energy, despite quasi-chaotic oscillatory dynamics of individual qubits. We also studied the interplay of bright and dark states for multiphoton excitation energies and determined the conditions for the formation of decoupled dark states.


\begin{acknowledgments}
This work has been supported in part by the Air Force Office for Scientific Research 
Grant No.~FA9550-21-1-0272, National Science Foundation Award No.~1936276, and Texas A\&M University through STRP, X-grant and T3-grant programs.

\end{acknowledgments}


\appendix

\section{Spatial distribution of the electric field in a plasmonic nanocavity}

In this section, we derive a representative example of the spatial distribution of the cavity field that we use in the numerical examples in this paper. We are interested in fields oscillating at optical frequencies confined a to three-dimensional plasmonic nanocavity. The optical wavelength is much larger than any characteristic length for the nanoscale confinement of the electric field; therefore, we can employ the quasistatic approximation to find the spatial distribution. Consider for definiteness the nanocavity created by a metallic sphere in (sub)nm vicinity to  the metallic substrate, as in strong coupling experiments with gold nanoparticles or in typical nanotip-enhanced optical experiments; see, e.g., \cite{chikkaraddy2016,park2016, pelton2018, park2019, may2021}. One can approximate the tip apex as a sphere, with the typical radius of $R \leq 10$ nm and variable distance to the substrate. 

To find the spatial distribution of the nanocavity field, we assume both the nanosphere and the substrate to be perfect conductors. The spatial field structure remains approximately the same in the presence of losses as long as the cavity quality factor $Q \gg 1$; in the opposite case, the notion of a cavity ceases to have any meaning. In typical experiments with metallic nanocavities  $10 \leq Q \leq 100$. Our problem, then, is that of solving the Laplace equation with Dirichlet boundary conditions on a sphere and a plane not intersecting the sphere.

\begin{figure}[htb]
    \includegraphics{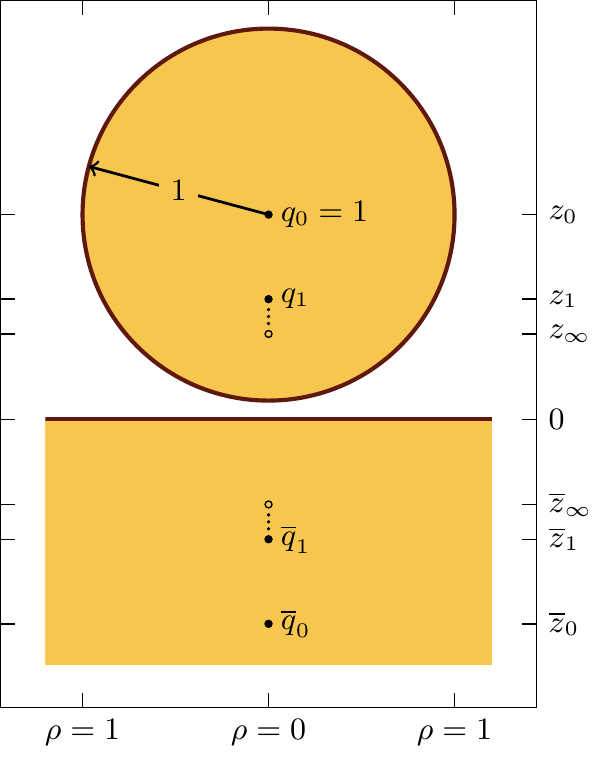}
    \caption{To-scale diagram of the first four image charges for $z_0 = 1.1$ with all scales normalized to the radius of the sphere. Also shown are $z_\infty \equiv \lim_{n\to\infty} z_n$ and $\overline z_\infty \equiv \lim_{n \to \infty} \overline z_n$ (we calculate these in the text). No image charges are placed in the sphere below $z_\infty$ and none are placed in the substrate above $\overline z_\infty$. The axes are the $z$ and $\rho$ of a cylindrical coordinate system.}
    \label{imageCharges}
\end{figure}

We work in a cylindrical system of coordinates with the origin on the plane and the cylindrical axis---the $z$-axis---intersecting the center of a sphere at $z = z_0 > 0$. The placement of the coordinate system is illustrated in Fig.~\ref{imageCharges}. We will normalize all spatial scales to the radius of the sphere. Since the sphere and the plane do not intersect, we have $z_0 > 1$.

We solve the problem using the method of images, as suggested in \cite{smythe1968}. This geometry requires the placement of an infinite number of point charges along the $z$-axis. Without loss of generality, we suppose the sphere to be at some positive potential and the plane to be at a potential of zero. Note that here we are interested only in the spatial field distribution; the amplitude is determined by the normalization condition \ref{nc-const} for the quantized field mode. 

First, we place an image charge $q_0$ at $z_0$---the center of the sphere; this raises the sphere to the desired nonzero potential. But $q_0$ breaks the boundary condition for the plane; the plane is distorted by $q_0$ to some nonzero, nonuniform potential. To restore the plane to ground, we place another image charge $\overline q_0 = - q_0$ at $\overline z_0 = - z_0$ inside the half-space---this is the reflection of $q_0$ in the plane. But now the boundary condition for the sphere is not satisfied. Typically, when it is introduced in elementary texts on electricity and magnetism (e.g., \cite{griffiths2017,jackson1999,landau1984}), correcting the distortion on the plane by $q_0$ (a plane and a point charge) is the first problem solved via the method of images and correcting the distortion on the sphere by $\overline q_0$ (a sphere and a point charge) is the second. To cancel the effect of $\overline q_0$ on the sphere, we place $q_1$ at $z_1$ such that $(z_0 - z_1)(z_0 - \overline z_0) = 1$ and $q_1 / \overline q_0 = - [(z_0 - z_1)/(z_0 - \overline z_n)]^{1/2}$. But now $q_1$ distorts the plane; so, we place $\overline q_1 = -q_1$ at $\overline z_1 = -z_1$, etc. The distortion of the plane by each $q_n$ is canceled by $\overline q_n$, the reflection of $q_n$ in the plane; the distortion of this $\overline q_n$ on the sphere is canceled by $q_{n+1}$, the reflection of $\overline q_n$ in the sphere. The first four image charges are depicted in Fig.~\ref{imageCharges}.

In the following section, we write a set of coupled difference equations (or recursion relations) for the image charges and their positions on the $z$-axis; we solve these equations to obtain closed form expressions for $q_n$ and $z_n$ in terms of the initial conditions $q_0$ and $z_0$; then, we write the field on the metallic substrate---the location of the quantum emitters---as an infinite series where each term is the contribution from $q_n$ and its reflection in the plane $\overline q_n$.

\subsection{Series solution via difference equations}

We set $q_0 = 1$, since the field amplitude is determined by normalization as already stated.
Then we have
\begin{align}
	\overline q_n &= -q_n,
	\label{chargePlaneReflect}
	\\
	\overline z_n &= - z_n,
	\label{displacementsPlaneReflect}
	\\
	( z_0 - z_{n+1} ) ( z_0 - \overline z_n ) &= 1,
	\label{displacementsWithOl}
	\\
	\frac {q_{n+1}} {\overline q_n} &= - \left( \frac {z_0 - z_{n+1}} {z_0 - \overline z_n} \right)^{1/2}
	\nonumber
	\\
	&= - \frac{1}{z_0 - \overline z_n},
	\label{chargeAndDisplacement}
\end{align}
where the second line of Eq.~(\ref{chargeAndDisplacement}) follows from Eq.~(\ref{displacementsWithOl}) and the fact that $1/(z_0 - \overline z_n) > 0$. Decoupled and with the $\overline q_n$s and $\overline z_n$s eliminated, Eqs.~(\ref{chargePlaneReflect}--\ref{chargeAndDisplacement}) are
\begin{align}
    (z_0 - z_{n+1})(z_0 + z_n) &= 1,
    \label{displacementsDiff}
    \\
    \frac 1 {q_n} + \frac 1 {q_{n+2}} &= \frac {2 z_0} {q_{n+1}}.
    \label{chargesDiff}
\end{align}

\eq{chargesDiff} is solved in \cite{smythe1968} but \eq{displacementsDiff} is not; we present solutions to both equations. The solution we present to \eq{chargesDiff} is similar to the solution in \cite{smythe1968}.

Eq.~(\ref{chargesDiff}) is a second-order, linear difference equation in $1/q_n$; furthermore, the zeroth ($1/q_n$) and second ($1/q_{n+2}$) terms are both multiplied by the same coefficient, namely, $1$. The solutions to this kind of equation are nicely expressed in terms of hyperbolic functions; this is due to the following two identities for hyperbolic functions:
\begin{align}
    \sinh{\vartheta n} + \sinh{\vartheta(n+2)} &= 2 \cosh {\vartheta} \sinh{\vartheta (n+1)},
    \nonumber
    \\
    \cosh{\vartheta n} + \cosh{\vartheta(n+2)} &= 2 \cosh {\vartheta} \cosh{\vartheta (n+1)}.
    \label{genSolHyperb}
\end{align}
Since $\sinh$ and $\cosh$ are linearly independent, Eq.~(\ref{genSolHyperb}) implies that
\begin{align}
    1/q_n = A \sinh{\alpha n} + B \cosh{\alpha n},
    \label{chargesGenSol}
\end{align}
where $\alpha$ defined by 
\begin{align}
    \cosh \alpha = z_0
\end{align}
is the general solution to Eq.~(\ref{chargesDiff}). The constants $A$ and $B$ can be determined from the given initial conditions $q_0$ and $z_0$. We use Eq.~(\ref{chargeAndDisplacement}) to find that $1/q_1 = 2 z_0 = 2 \cosh{\alpha}$; thus, $A$ and $B$ are determined by the system
\begin{align}
    1/q_0 = 1 &= A,
    \nonumber
    \\
    1/q_1 = 2 \cosh\alpha &= A \cosh\alpha + B \sinh\alpha.
    \label{chargesSys}
\end{align}
Eq.~(\ref{chargesSys}) is solved by $A = 1$ and $B = 1/\tanh\alpha$; so, the image charges are given by
\begin{align}
    q_n = \frac{\sinh \alpha}{\sinh{\alpha(n+1)}}.
    \label{charges}
\end{align}
In writing \eq{charges}, we have used the identity
\begin{align}
   \sinh \vartheta \cosh \varphi + \cosh \vartheta \sinh \varphi = \sinh{(\vartheta + \varphi)}
   \label{hyperbSum}
\end{align}
to simplify the expression obtained from substituting the values of $A$ and $B$ found from solving \eq{chargesSys} into \eq{chargesGenSol}.

We have obtained a closed-form expression for $q_n$; now, we turn our attention toward doing the same for $z_n$. While \eq{displacementsDiff} is nonlinear, it is first-order and rational; furthermore, it is of a form such that it can be reduced to a linear second-order difference equation via a simple nonlinear change of variable---this method is detailed in \cite{brand1955}. We rearrange \eq{displacementsDiff} by solving for $z_{n+1}$ and adding $z_0 = \cosh{\alpha}$ to both sides:
\begin{align}
    z_{n+1} + \cosh{\alpha} &= 2 \cosh{\alpha} - \frac 1 {z_n + \cosh\alpha}.
    \label{zeta}
\end{align}
We write \eq{zeta} in terms of the new variable $\xi_n$ where the $\xi_n$s are defined by $z_n + \cosh\alpha = \xi_{n+1}/\xi_{n}$; this leads to
\begin{align}
    \xi_{n+2} + \xi_{n} = \left. 2 \cosh{\alpha} \right. \xi_{n+1}.
    \label{xi}
\end{align}
\eq{xi} is identical to \eq{chargesDiff}; so, \eq{xi} is also solved by \eq{chargesGenSol}, which implies
\begin{align}
    z_n + \cosh{\alpha} = \frac{\cosh{\alpha (n+1)} + C \sinh{\alpha (n+1)}}{\cosh{\alpha n} + C \sinh{\alpha n}}.
    \label{displacementsGenSol}
\end{align}

Note that \eq{displacementsGenSol} contains only one undetermined constant while \eq{chargesGenSol}---from which \eq{displacementsGenSol} is derived---contains two. This is due to the fact that \eq{displacementsGenSol} is the general solution to \eq{displacementsDiff}, which is first-order, while \eq{chargesGenSol} is second-order.

Applying the initial condition $z_0 = \cosh{\alpha}$ to \eq{displacementsGenSol} yields $C = 1/\tanh{\alpha}$ which leads to
\begin{align}
    z_n = \frac{\sinh{\alpha}}{\tanh{\alpha(n+1)}}.
    \label{displacements}
\end{align}
To obtain \eq{displacements}, we have again used \eq{hyperbSum} to simplify.

\begin{figure}[htb]
    \includegraphics{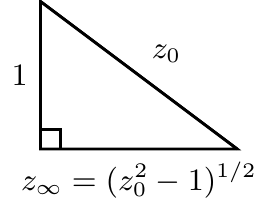}
    \caption{The existence of a triangle with sides of these lengths is ensured by the Pythagorean theorem. The triangle inequality applied to this triangle yields $1 + (z_0^2 - 1)^{1/2} > z_0$; so, $z_\infty > z_0 - 1$. See \eq{z_inf}.}
    \label{triangle}
\end{figure}

Using the expressions obtained for $q_n$ and $z_n$, we can write the field $\mathbf E$ as an infinite series; but first, we consider the asymptotic behaviors of $q_n$ and $z_n$ for large $n$. Since, for large $n$, $\sinh{\alpha (n+1)}$ behaves like $e^{|\alpha|n}$, $q_n$ rapidly approaches $0$. On the other hand, $z_n$ rapidly approaches the constant $\left| \sinh \alpha \right|$ since $\tanh {\alpha (n+1)}$ rapidly approaches $1$ if $\alpha > 0$ or $-1$ if $\alpha < 0$. Denote
\begin{align}
    z_\infty \equiv \lim_{n\to\infty} z_n = \left| \sinh{\alpha} \right| = (z_0^2 - 1)^{1/2}.
    \label{z_inf}
\end{align}
The last equality follows from the identity $\sinh{\arcosh x} = (x^2 - 1)^{1/2} \text{ which holds for all } x \text{ such that } |x| > 1$. Since $z_\infty = (z_0^2 -1)^{1/2}$, $z_\infty > z_0 - 1$ (i.e., the point $z = z_\infty$ on the $z$-axis is inside the ball) follows from the triangle inequality---see Fig.~\ref{triangle}; so, all image charges are placed inside one of the conductors ($z_n$ strictly decreases from $z_0$ as $n$ increases so $z_0 \geq z_n > z_\infty$ for all $n$), as expected.

\begin{figure}[htb]
    \includegraphics{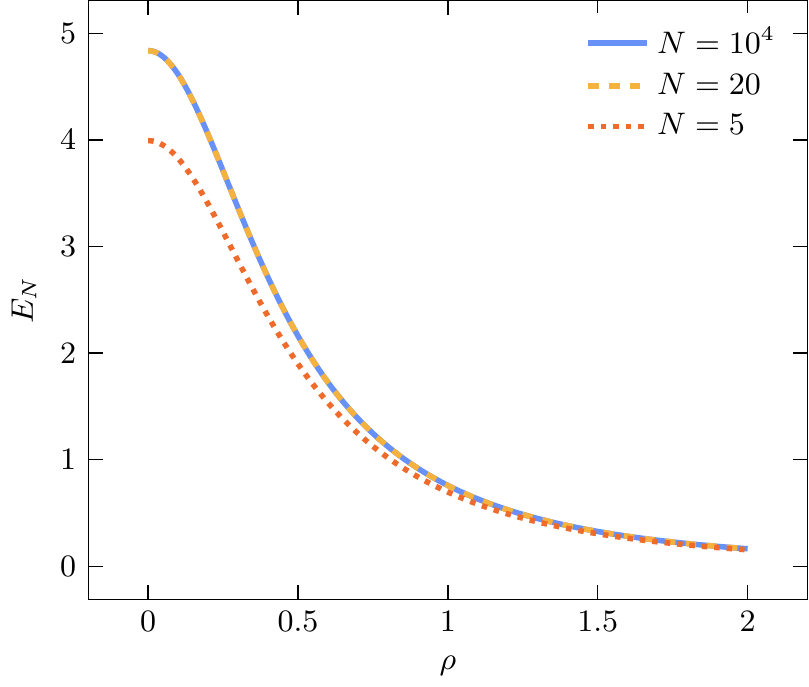}
    \caption{$E_N$ plotted against $\rho$ for $z_0 = 1.1$ and various values of $N$. $E_N$ is the field due to the first $2N$ image charges; it is the first $N$ terms in the series solution for the field---\eq{fieldSeries}. The lines corresponding to $E_20$ and $E_{10^4}$ are indistinguishable.}
    \label{fieldConvergence}
\end{figure}

We are interested in the field $\mathbf E$ in the $z = 0$ plane. In this plane we have 
\begin{align}
    \mathbf E &= \sum_{n=0}^\infty \left[ \frac{q_n (\boldsymbol\rho - \mathbf{\hat z} z_n)}{(\rho^2 + z_n^2)^{3/2}} - \frac{q_n(\boldsymbol\rho+\mathbf{\hat z} z_n)}{(\rho^2 + z_n^2)^{3/2}} \right]
    \nonumber
    \\
    &= - 2\mathbf{\hat z} \sum_{n=0}^\infty \frac{q_n z_n}{(\rho^2 + z_n^2)^{3/2}}.
    \label{fieldOfPoints}
\end{align}
Since the field in the plane, as computed with \eq{fieldOfPoints}, is purely in the $z$ direction, we will from now on write the magnitude of the field $E = -E_z$ instead of the field $\mathbf E$; also, since we are going to normalize the field, we drop the prefactor of 2 on the second line of \eq{fieldOfPoints}. Substituting Eqs.~(\ref{charges}) and (\ref{displacements}) into \eq{fieldOfPoints}, we arrive at
\begin{align}
    E_N \equiv \sinh^2 \alpha \sum_{n=1}^N \frac{\cosh{\alpha n}}{\sinh^2{\alpha n}} \bigg[ \rho^2 + \Big( \frac{R \sinh\alpha}{\tanh{\alpha n}} \Big)^2 \bigg]^{-3/2} \to E \text{ as } N \to \infty.
	\label{fieldSeries}
\end{align}
For large $n$, the $n$th term in \eq{fieldSeries} is proportional to $e^{-|\alpha| n}$; the series converges rapidly---more rapidly for larger values of $|\alpha|$, that is, for larger values of $z_0$ ($\cosh x$ is increasing on $x\in(0,\infty)$). Consider Fig.~\ref{fieldConvergence} which illustrates convergence of the series for the case $z_0 = 1.1$; $E_N$ does not change appreciably between $N = 20$ and $N = 10^4$. Furthermore, for $z_0 \geq 2$ it is enough to have $N = 3$.

\subsection{Two approximations}

\eq{fieldSeries} is straightforward to use in numerical simulation but is unwieldy for the analytic derivation of the quantum dynamics in the main text. We consider two physically motivated approximations; we call one the \textit{point-charge approximation} and the other the \textit{line-charge approximation}.

We define the point-charge approximation
\begin{align}
    E_N^ p \equiv \frac{Q_N Z_N}{(\rho^2 + Z_N^2)^{3/2}}
    \label{pointApprox}
\end{align}
where $Q_N \equiv \sum_0^{N-1} q_n$ and $Z_N \equiv (1/Q_N) \sum_0^{N-1} z_n q_n$. The point-charge approximation is the field due to a real dipole composed of $Q_N$ at $z = Z_N$ and $-Q_N$ at $-Z_N$. By ``real dipole'' we mean two point charges with charges of opposite sign but equal magnitude separated by some finite distance. Unlike the field due to a dipole vector located at the origin, the field of this real dipole does not diverge for small as $\rho \to 0$. $Q_N$ is just the total sum of charges. $Z_N$ is the average of the displacements of the charges weighted according to the magnitude of the charges, i.e., it is the position of the $q_n$'s center of mass but with charge instead of mass. The point-charge approximation is the field due to the point charge which most closely resembles the infinite series of image charges above the substrate and that most-closely-resembling charge's reflection in the plane.

The line charge approximation is 
\begin{align}
    E_N^ l \equiv \frac {Q_N} {z_0 - \sqrt{z_0^2 - 1}} \left( \frac 1 {\sqrt{\rho^2 + z_0^2 - 1}} - \frac 1 {\sqrt{\rho^2 + z_0^2}} \right)
    \label{lineApprox}
\end{align}
where $Q_N$ is the same as in the point-charge approximation. The line-charge approximation is the field due to a total charge of $Q_N$ distributed uniformly over the line between $z_n$ and $z_\infty$ and the reflection of this object in the plane. We chose a uniform charge distribution not because the discrete distribution of image charges are well approximated by uniform continuous distribution---it is not---but because it is simple and because it leads to an integrand with a nice antiderivative. The line-charge approximation is the most straightforward way to include the fact that the $q_n$'s are extended in the $z$-direction.

While the point-charge approximation is a single term, its dependence on $z_0$---through $Z_N$ and $Q_N$---is not expressed in closed form. On the other hand, while the line-charge approximation is two terms, its dependence on $z_0$ is simpler; it still contains $Q_N$ but does not contain $Z_N$. We will see that the line-charge approximation is also more accurate for $z_0 \sim 1$ which is our main interest.


\begin{figure}[htb]
    \includegraphics{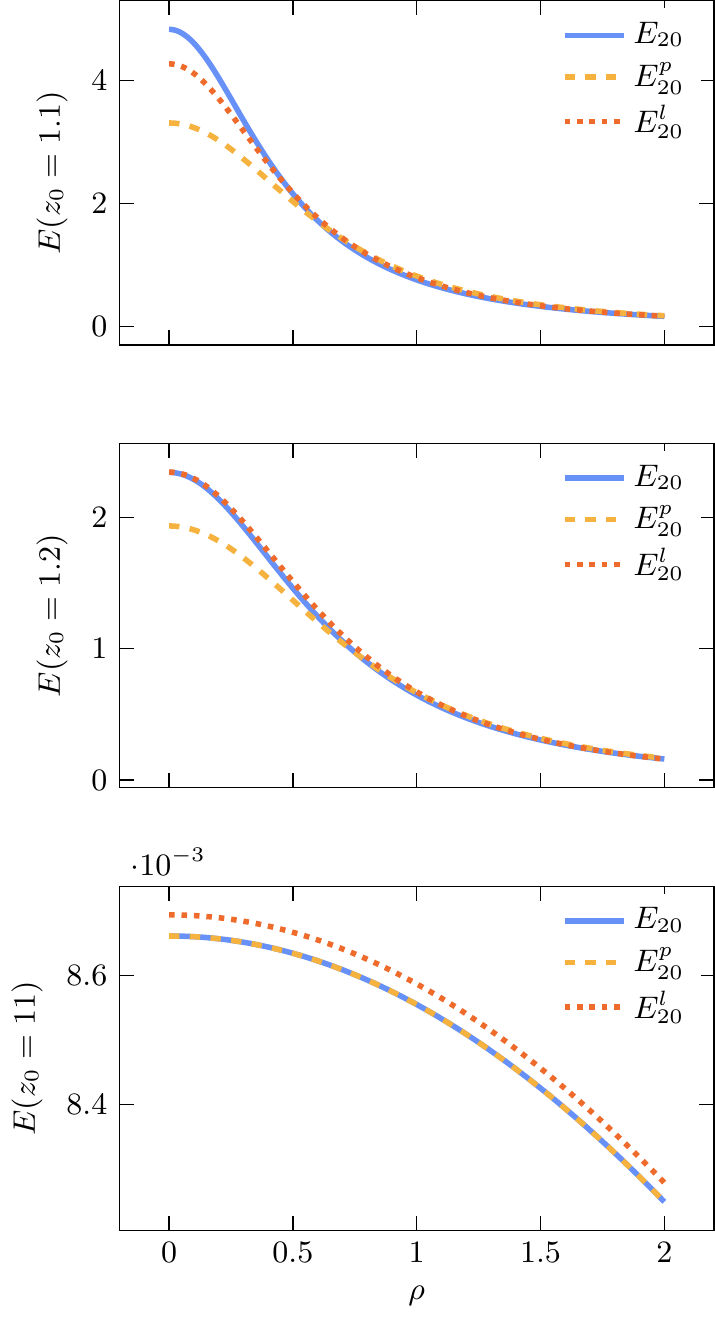}
    \caption{Three plots of $E_{20}$ (the series solution---\eq{fieldSeries}), $E_{20}^p$ (the point-charge approximation---\eq{pointApprox}) and $E_{20}^l$ (the line-charge approximation---\eq{lineApprox}) against $\rho$ for three values of $z_0$, namely, 1.1 (top), 1.2 (middle) and 11 (bottom). In the $z_0 = 11$ plot, the lines corresponding to $E_{20}$ and $E^p_{20}$ are indistinguishable. The numerical examples in the main text make use of the line approximation with $z_0 = 1.2$.}
    \label{zDependence}
\end{figure}

\begin{figure}[htb]
    \includegraphics{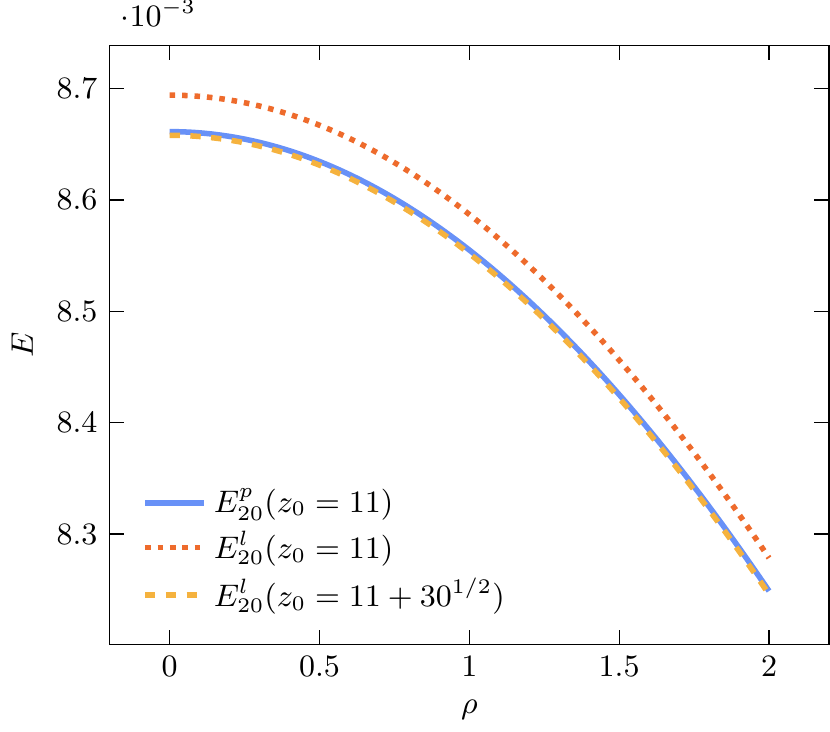}
    \caption{Plot of the point-charge (\eq{pointApprox}) and line-charge (\eq{lineApprox}) approximations evaluated at $z_0 = 11$ as well as the line-charge approximation evaluated at $11 + [11 - (11^2 - 1)^{1/2}]/2 = 11 + 30^{1/2}$ against $\rho$. The lines correspond to $E^p_{20}(z_0 = 11)$ and $E^l_{20}(z_0 = 11+30^{1/2})$ are nearly indistinguishable.}
    \label{asympBehavior}
\end{figure}


Fig.~\ref{zDependence} illustrates the accuracy of the point and line-charge approximations at three values of $z_0$. We evaluate the accuracy of the approximations by comparing them to $E_{20}$ (since $N = 20$ is enough terms for the series to converge at $z_0 = 1.1$---see Fig.~\ref{fieldConvergence}, it should also be enough for $z_0 = 1.2$ and $z_0 = 11$). The line-charge approximation is more accurate than the point-charge approximation for $z_0 \sim 1$. For $z_0 < \approx 1.2$ the line-charge approximation underestimates the field at small $\rho$ and for $z_0 > \approx 1.2$ the line-charge approximation overestimates the field at small $\rho$. For large $z_0$ (e.g., the $z_0 = 11$ plot in Fig.~\ref{zDependence}), while the point-charge approximation has converged to the true field, the line-charge approximation hovers above the true field, that is, the line-charge approximation overestimates the field by about the same amount for all $\rho$. The origin of this hovering behavior becomes apparent when we consider the point-charge approximation of the line-charge approximation, i.e., an approximation of an approximation (and the point-charge approximation is appropriate for any distant-from-the-origin and localized charge distribution mirrored in the plane). The best fit point charge to the line of charge involved in the line-charge approximation is a charge located at the center of the line, that is, at $z = z_0 - (z_0 - z_\infty)/2$; however, this is slightly too close to the origin---for large $z_0$, the exact field is best approximated by a point charge at $z = z_0$. Fig.~\ref{asympBehavior} corroborates this argument; it shows that, for $z_0 = 11$, when the substitution $z_0 \to z_0 + (z_0 - z_\infty)/2$ is made, the point and line-charge approximations agree.


\section{The limit of a continuous spectrum of transition frequencies}

Consider a large enough sample of qubits with a dense enough distribution of transition frequencies, so that the continuous distribution limit in Eq.~(\ref{C10-5})  is justified. This is possible when 
\begin{equation}
| \Omega_{Rj}| \gg   \Delta_j - \Delta_{j\pm 1}. 
 \nonumber
\end{equation}
In the opposite limit the field is mostly coupled to one qubit closest to resonance. 

In the continuous limit we replace $\Delta_j \Rightarrow \Delta$ and introduce the density of states $g(\Delta)$ as
\begin{equation}
\sum_{j=k}^{k+p} O_j = \int_{\Delta_k}^{\Delta_{k+p}} O({\Delta}) g(\Delta) d\Delta, 
 \nonumber
\end{equation}
where $O_j$ is a sequence of discrete values of a given function. 

Then Eq.~(\ref{C10-5}) is transformed as
\begin{equation}
\dot{C}_{10} + \frac{\mu}{2} C_{10} = i \int_{-\infty}^{\infty} e^{-i\Delta t} \Omega_{R\Delta}^* C_{0\Delta}(0)  g(\Delta) d\Delta - \int_0^t \left[ \int_{-\infty}^{\infty} e^{i\Delta (\tau - t)} |\Omega_{R\Delta}|^2  g(\Delta) d\Delta \right] C_{10}(\tau)  d\tau.
 \label{C10-5cont}
\end{equation}

It is convenient to parameterize the density of states  $g(\Delta)$ and $\Omega_{R\Delta}$ as
\begin{equation}
g(\Delta) = \frac{N}{2 \Delta_m} f(\Delta), \;  \Omega_{R\Delta} = \frac{\Omega_N}{\sqrt{N}} \rho(\Delta),
 \nonumber
\end{equation}
where $2\Delta_m$ is the width of the distribution of frequency detunings. With this parameterization $\int_{-\infty}^{\infty}  f(\Delta) d\Delta = 2 \Delta_m$. 
As a result, Eq.~(\ref{C10-5cont}) takes a form convenient for applying the Laplace transform:
\begin{equation}
\dot{C}_{10} + \frac{\mu}{2} C_{10} = i \frac{\sqrt{N} \Omega_N}{2 \Delta_m} \tilde{F}(t)- \frac{|\Omega_N|^2}{2 \Delta_m} \int_0^t \tilde{D}(t-\tau) C_{10}(\tau)  d\tau,
 \label{C10-7cont}
\end{equation}
where 
\begin{equation}
\tilde{F}(t)=  \int_{-\infty}^{\infty} F_{\Delta} e^{-i\Delta t} d\Delta; \; F_{\Delta} =  C_{0\Delta}(0) \rho^*(\Delta) f(\Delta);  
\nonumber
\end{equation}
 \begin{equation}
\tilde{D}(t) =  \int_{-\infty}^{\infty} D_{\Delta} e^{-i\Delta t} d\Delta; \;  D_{\Delta} = |\rho(\Delta)|^2 f(\Delta).
\nonumber
\end{equation}
Since $\int_{-\infty}^{\infty}  |\Omega_{R\Delta}|^2  g(\Delta) d\Delta = \sum_{j=1}^N |\Omega_{Rj}|^2 = \Omega_N^2$, one can show that 
$\int_{-\infty}^{\infty}  D_{\Delta} d\Delta = 2 \Delta_m$. 

Applying Laplace transform to Eq.~(\ref{C10-7cont}) gives
\begin{equation}
pC_p - C_{10}(0) + \frac{\mu}{2} C_p = i \frac{\sqrt{N} \Omega_N}{2 \Delta_m} \tilde{F}_p- \frac{|\Omega_N|^2}{2 \Delta_m} C_p \tilde{D}_p,
 \label{C10-8cont}
\end{equation}
where 
\begin{equation}
\tilde{F}_p=  \int_{0}^{\infty} \tilde{F}(t)  e^{-p t} dt  = \int_{-\infty}^{\infty} \frac{F_{\Delta}}{i\Delta + p}  d\Delta, 
\nonumber
\end{equation}
\begin{equation}
\tilde{D}_p=  \int_{0}^{\infty} \tilde{D}(t)  e^{-p t} dt  = \int_{-\infty}^{\infty} \frac{D_{\Delta}}{i\Delta + p}  d\Delta.  
\nonumber
\end{equation}
Solving Eq.~(\ref{C10-8cont}) gives
\begin{equation}
C_{10}(t)  = \frac{1}{2 \pi i}  \int_{x-i\infty}^{x+i\infty} \frac{C_{10}(0) + i \frac{\sqrt{N} \Omega_N}{2 \Delta_m} \int_{-\infty}^{\infty} \frac{F_{\Delta} d\Delta}{i\Delta + p} }{p + \frac{\mu}{2} +  \frac{|\Omega_N|^2}{2 \Delta_m}  \int_{-\infty}^{\infty} \frac{D_{\Delta} d\Delta}{i\Delta + p} } e^{pt} dp, 
 \label{C10-9cont}
\end{equation}
where, as usual, the analytic continuation of the complex $p$-plane to the region Re$[p] \leq 0$ corresponds to the counterclockwise integration path around the poles in the integrals $  \int_{-\infty}^{\infty} \frac{(...) d\Delta}{i\Delta + p}$. 

The poles of an integrand in Eq.~(\ref{C10-9cont}) are determined by  
\begin{equation}
p + \frac{\mu}{2} + \frac{|\Omega_N|^2}{2 \Delta_m} \int_{-\infty}^{\infty} \frac{D_{\Delta}}{i\Delta + p}  d\Delta = 0.   
\label{poles-cont} 
\end{equation}

In our system the inhomogeneous broadening is much greater than the decay rate of the cavity field, $\Delta_m \gg \mu$; see Eq.~(\ref{dominates}). In the strong coupling regime, the Rabi frequency is also much greater than the cavity decay rate, $\Omega_N \gg \mu$. The value of the ratio between $\Delta_m$ and $\Omega_N$ determines two distinct dynamic regimes. 


\subsection{Strong inhomogeneous broadening}

In this case 
 \begin{equation}
 \frac{|\Omega_N|^2}{2 \Delta_m^2} \ll 1 \ll   \frac{N |\Omega_N|^2}{4 \Delta_m^2},
\label{strong_ib}
\end{equation}
where the second inequality is due to the limit of a continuous spectrum: the typical value of the Rabi frequency $\langle \Omega_R \rangle \sim \frac{\Omega_N}{\sqrt{N}}$ should exceed the distance between discrete spectral lines $2\Delta_m/N$. The first inequality ensures that near the pole the value of $|p| \sim {\rm max}\left[ \mu, \frac{\Omega_N^2}{\Delta_m} \right] \ll \Delta_m$. In this case, taking into account correct direction of the integration path around the pole in Eq.~(\ref{C10-9cont}), we obtain a standard expression:
 \begin{equation}
 \frac{1}{i\Delta + p}  \Rightarrow \pi \delta(\Delta) - i \frac{\mathcal{P}}{\Delta},  
\nonumber
\end{equation}
where $\mathcal{P}$ is principal value of the integral. The resulting solution of Eq.~(\ref{poles-cont}) is 
 \begin{equation}
p_0 = - \frac{\mu}{2} - \frac{\pi |\Omega_N|^2}{2 \Delta_m} D_{\Delta=0} + i \frac{|\Omega_N|^2}{2 \Delta_m}\int_{-\infty}^{\infty} \frac{\mathcal{P}}{\Delta} D_{\Delta}  d\Delta.  
\label{poles2} 
\end{equation}

It is easy to show that the expression $\frac{\pi |\Omega_N|^2}{2 \Delta_m} D_{\Delta=0} $ is {\it exactly} the probability of transition per unit time from the state $ | 1 \rangle \Pi_{j=1}^N | 0_j \rangle$ into the continuous spectrum of states of excited qubits calculated with Fermi's Golden Rule.

  The time evolution of $C_{10}(t)$ becomes 
  \begin{equation}
C_{10}(t) \approx  \left[ C_{10}(0)  + \frac{\sqrt{N} \Omega_N}{2 \Delta_m}   \left(  i \pi F_{\Delta = 0} + \int_{-\infty}^{\infty}  \frac{\mathcal{P}}{\Delta} F_{\Delta}  d\Delta \right) \right] e^{p_0 t}.  
 \label{C10-10}
\end{equation}
The second term in the brackets on the rhs of Eqs.~(\ref{C10-10}) is due to the dynamics at short times $t < \Delta_m^{-1} \ll \mu^{-1}$, i.e., before the exponential decay kicks in. If qubits are not initially excited, this term is exactly zero. Furthermore, it can be neglected if the initial  probability of finding the photon mode excited, $P_{ph} = |C_{10}(0)|^2$, is at least as large as the initial excitation of the qubits,  $P_{qub} = \sum_{j=1}^N |C_{0j}(0)|^2$, whereas the distribution of excitation probabilities of individual qubits is  ``uniform'': $|F_{\Delta}| \sim   |C_{0j}(0)| \sim \frac{1}{\sqrt{N}}  |C_{10}(0)|$. 

It is important to keep in mind that despite Eq.~(\ref{dominates}), dissipation of the cavity field can be faster than the energy transfer to resonant qubits, as long as 
\begin{equation}
 1 \gg \frac{\mu}{4 \Delta_m} > \pi \frac{\Omega_N^2}{4 \Delta_m^2}. 
\nonumber
\end{equation}


\subsection{``Weak'' inhomogeneous broadening}

Now consider a relatively narrow frequency spectrum, when 
\begin{equation}
\frac{\Omega_N^2}{2 \Delta_m^2} \gg 1, 
\nonumber
\end{equation}
while still $\Delta_m \gg \mu$. 
In this case the transition to continuous spectrum is always valid and the roots of Eq.~(\ref{poles-cont}) satisfy $|p| \sim \Omega_N \gg \Delta_m$. 
Since the typical width of the spectrum is $2\Delta_m$, we always have $D_{|\Delta| \sim \Omega_N \gg \Delta_m} \ll 1$, or even $D_{|\Delta| > \Delta_m} = 0$ for a limited spread of transition frequencies. Keeping only the leading nonzero terms with respect to a small parameter $\frac{\Delta_m}{\Omega_N}$ and using $\int_{-\infty}^{\infty} D_{\Delta}  d\Delta = 2 \Delta_m$, Eq.~(\ref{poles-cont}) can be transformed to 
 \begin{equation}
p^2 + \Omega_N^2 \approx - \left[ \frac{\mu}{2} +\frac{\Omega_N^2}{2 \Delta_m} \left( D_{\Delta={\rm Im}[p]} - \frac{i}{p^2} \int_{-\infty}^{\infty} \Delta D_{\Delta}  d\Delta + \frac{1}{p^3} \int_{-\infty}^{\infty} \Delta^2 D_{\Delta}  d\Delta \right) \right] p,   
\label{poles3} 
\end{equation}
which has the solution
 \begin{equation}
p_0 = \pm i ( \Omega_N - \delta \Omega_s) - i \delta \Omega_{as} - \kappa_{\pm} + o\left(\frac{(\delta \Omega_{s,as})^2}{\Omega_N}, \frac{\kappa_{\pm}^2}{\Omega_N} \right), 
\label{poles4} 
\end{equation}
where
 \begin{equation}
\delta \Omega_s = \frac{1}{4 \Delta_m \Omega_N} \int_{-\infty}^{\infty} \Delta^2 D_{\Delta}  d\Delta , \; \delta \Omega_{as}  = -\frac{1}{2 \Delta_m} \int_{-\infty}^{\infty} \Delta D_{\Delta}  d\Delta, \; \kappa_{\pm} = \frac{\mu}{4} +\frac{\pi \Omega_N^2}{4 \Delta_m} D_{\Delta=\pm \Omega_N}. 
\nonumber
\end{equation}

In particular, for Gaussian distribution $D_{\Delta} = \frac{2}{\sqrt{\pi}} e^{-\Delta^2/\Delta_m^2}$,
\begin{equation}
\delta \Omega_s = \frac{ \Delta_m^2}{ 2 \sqrt{\pi} \Omega_N}, \; \delta \Omega_{as}  =0, \; \kappa_+ = \kappa_- = \frac{\mu}{4} +\frac{\sqrt{\pi}  \Omega_N^2}{2 \Delta_m} e^{-\Omega_N^2/\Delta_m^2}. 
\nonumber
\end{equation}
Note that the contribution to photon absorption $\kappa_{\pm}$ originated from light-qubit coupling (the second term) cannot be expanded in powers of a small parameter $\frac{\Delta_m}{\Omega_N}$. 

Comparing this solution with the one obtained without any inhomogeneous broadening, it is easy to see that the frequency shift due to inhomogeneous broadening, $\sim \frac{\Delta_m^2}{\Omega_N}$, is always greater than the one due to finite cavity field decay,  $\sim \frac{\mu^2}{\Omega_N}$, as long as Eq.~(\ref{dominates}) is satisfied. At the same time, photon absorption $\kappa_{\pm}$ is dominated by the cavity field dissipation $\mu$. This is obvious when the spread of frequencies is limited and $D_{\Delta=\pm \Omega_N} = 0$, but it remains true also for a Gaussian distribution $D_{\Delta} $ as long as 
\begin{equation}
1 \gg  \frac{\mu}{4 \Delta_m} > \frac{\sqrt{\pi}}{2} \frac{  \Omega_N^2}{ \Delta_m^2} e^{-\Omega_N^2/\Delta_m^2}. 
\nonumber
\end{equation}

To get simpler algebra, let's consider a symmetric distribution $D_{\Delta} $ when  $\kappa_+ = \kappa_-$ and $\delta \Omega_{as}  = 0$. A general case leads to more cumbersome expressions but the same qualitative result. Neglecting the terms of the order of  $   \frac{\int_{-\infty}^{\infty} \Delta F_{\Delta} d\Delta}{\Omega_N \int_{-\infty}^{\infty} F_{\Delta} d\Delta}\sim \frac{\Delta_m}{\Omega_N}$, $   \frac{\int_{-\infty}^{\infty} \Delta^2 F_{\Delta} d\Delta}{\Omega_N^2 \int_{-\infty}^{\infty} F_{\Delta} d\Delta}\sim \frac{\Delta_m^2}{\Omega_N^2}$, $\frac{\mu}{\Omega_N}$ etc., we obtain a result similar to the one for identical qubits without detunings. Indeed, in this case Eq.~(\ref{C10-9cont}) gives the following the solution for $C_{10}(t)$: 
 \begin{equation}
C_{10}(t) \approx  \left( C_{10}(0) \cos\left[ ( \Omega_N - \delta \Omega_s)t\right] + i \frac{F(0)}{\Omega_N} \sin\left[ ( \Omega_N - \delta \Omega_s)t\right] \right) e^{-\kappa \tau},
\label{C10-11cont}
\end{equation}
where we used 
 \begin{equation}
 \frac{\sqrt{N} \Omega_N}{2 \Delta_m}  \int_{-\infty}^{\infty} F_{\Delta}  d\Delta = \sum_{j=1}^N \Omega_{Rj}^* C_{0j}(0) = F(0).  
 \nonumber 
\end{equation}
Substituting  Eq.~(\ref{C10-11cont}) into Eq.~(\ref{C0j-5}) yields
 \begin{equation}
C_{0j}(t) =  C_{0j}(0) + i \Omega_{Rj} \int_0^t \left( C_{10}(0) \cos\left[ ( \Omega_N - \delta \Omega_s)t\right] + i \frac{F(0)}{\Omega_N} \sin\left[ ( \Omega_N - \delta \Omega_s)t\right] \right) e^{(i \Delta_j -\kappa) \tau} d\tau. 
\label{C0j-11}
\end{equation}



\begin{thebibliography}{99}


\bibitem{thorma2015} P. T\"{o}rma and W. L. Barnes, Strong coupling between
surface plasmon polaritons and emitters: a review, Rep. Prog. Phys. 78,
013901 (2015).



\bibitem{lodahl2015} P. Lodahl, S. Mahmoodian, and S. Stobbe, Interfacing
single photons and single quantum dots with photonic nanostructures, Rev.
Mod. Phys. 87, 347 (2015).

\bibitem{degen2017} C. L. Degen, F. Reinhard, and P. Cappellaro, Quantum
sensing, Rev. Mod. Phys. 89, 035002 (2017).

\bibitem{dovzhenko2018} D. S. Dovzhenko,  S. V. Ryabchuk, Yu. P. Rakovich,  and I. R. Nabiev, Light-matter interaction in the strong coupling
regime: configurations, conditions, and applications, Nanoscale 10, 3589 (2018). 


\bibitem{bitton2019} O. Bitton, S. N. Gupta, and G. Haran, Quantum dot
plasmonics: from weak to strong coupling, Nanophotonics 8, 559 (2019).

\bibitem{tavis1968} M. Tavis and F. W. Cummings, Exact solution for an N-molecule-radiation-field Hamiltonian, Phys. Rev. 170, 379-384 (1968).  

\bibitem{shammah2018} N. Shammah, S. Ahmed, N. Lambert, S. De Liberato, and F. Nori, Open quantum systems with local and collective incoherent processes: Efficient numerical simulations using permutational invariance, Phys. Rev. A 98, 063815 (2018). 

\bibitem{rose2017}   B. C. Rose, A. M. Tyryshkin, H. Riemann, N. V. Abrosimov, P. Becker, H.-J. Pohl, 
M. L.W. Thewalt, K. M. Itoh, and S. A. Lyon, Coherent Rabi Dynamics of a Superradiant Spin Ensemble in a Microwave Cavity, Phys. Rev. X 7, 031002 (2017). 

\bibitem{laucht2010} A. Laucht, J. M. Villas-Boas, S. Stobbe,3 N. Hauke, F. Hofbauer, G. Bohm, P. Lodahl, M.-C. Amann, M. Kaniber,
and J. J. Finley, Mutual coupling of two semiconductor quantum dots via an optical nanocavity, Phys. Rev. B 82, 075305 (2010). 

\bibitem{gray2015} M. Otten, R. A. Shah, N. F. Scherer, M. Min, M. Pelton,
and S. K. Gray, Entanglement of two, three, or four plasmonically coupled
quantum dots, Phys. Rev. B 92, 125432 (2015).

\bibitem{gray2016} M. Otten, J. Larson, M. Min, S. M. Wild, M. Pelton, and
S. K. Gray, Origins and optimization of entanglement in plasmonically
coupled quantum dots, Phys. Rev. A 94, 022312 (2016).







\bibitem{zoller1997} P. Zoller and C. W. Gardiner. Quantum Noise in Quantum Optics: the Stochastic Schr\"{o}dinger Equation. Lecture Notes for the Les Houches Summer School LXIII on Quantum Fluctuations in July 1995, Elsevier Science Publishers B.V. 1997, edited by E. Giacobino and S. Reynaud;  arXiv:quant-ph/9702030v1. 

\bibitem{Plenio1998} M. B. Plenio and P. L. Knight, The quantum-jump
approach to dissipative dynamics in quantum optics, Rev. Mod. Phys. 70, 101
(1998).


\bibitem{gisin1992} N. Gisin and I. C. Percival, The quantum-state diffusion model applied to open systems, J. of Phys. A: Mathematical and General 25 , 5677-5691 (1992).

\bibitem{diosi1998} L. Diosi, N. Gisin, and W. T. Strunz, Non-Markovian quantum state diffusion, Phys. Rev. A 58,  1699-1712 (1998).

\bibitem{cohen1993} C. Cohen-Tannoudji, B.o Zambon and E. Arimondo, Quantum-jump approach to dissipative processes: application to amplification without inversion, J. Opt. Soc. Am. B 10,  2107-2120 (1993). 

\bibitem{molmer1993} K. Molmer, Y. Castin and J. Dalibard, Monte Carlo wave-function method in quantum optics, J. Opt. Soc. Am. B 10, 524-538 (1993).

\bibitem{gisin1992-2} N. Gisin and I. C. Percival, Wave-function approach to dissipative processes: are there quantum jumps? Phys. Lett. A 167,  315-318 (1992). 



\bibitem{tokman2020} M. Tokman, M. Erukhimova, Y. Wang, Q. Chen, and A.
Belyanin, Generation and dynamics of entangled fermion-photon-phonon states
in nanocavities, Nanophotonics 10, 491 (2021). 

\bibitem{chen2021} Q. Chen, Y. Wang, S. Almutairi, M. Erukhimova, M. Tokman,
and A. Belyanin. Dynamics and control of entangled electron-photon states in
nanophotonic systems with time-variable parameters, Phys. Rev. A 103, 013708
(2021). 



\bibitem{todorov2010} Y. Todorov, A. M. Andrews, R. Colombelli, S. De
Liberato, C. Ciuti, P. Klang, G. Strasser, and C. Sirtori, Ultrastrong
Light-Matter Coupling Regime with Polariton Dots, Phys. Rev. Lett. 105,
196402 (2010).

\bibitem{forndiaz2017} P. Forn-Diaz, J. J. Garcia-Ripoll, B. Peropadre,
J.-L. Orgiazzi, M. A. Yurtalan, R. Belyansky, C. M.Wilson, and A. Lupascu,
Ultrastrong coupling of a single artificial atom to an electromagnetic
continuum in the nonperturbative regime, Nat. Phys. 13, 39 (2017).

\bibitem{kono2019} P. Forn-Diaz, L. Lamata, E. Rico, J. Kono, and E. Solano,
Ultrastrong coupling regimes of light-matter interaction, Rev. Mod. Phys.
91, 025005 (2019).





\bibitem{chikkaraddy2016} R. Chikkaraddy, B. de Nijs, F. Benz, S. J. Barrow,
O. A. Scherman, E. Rosta, A. Demetriadou, P. Fox, O. Hess, and J. J.
Baumberg, Single-molecule strong coupling at room temperature in plasmonic
nanocavities, Nature 535, 127 (2016).

\bibitem{benz2016} F. Benz, M. K. Schmidt, A. Dreismann, R. Chikkaraddy, Y.
Zhang, A. Demetriadou, C. Carnegie, H. Ohadi, B. de Nijs, R. Esteban, J.
Aizpurua, and J. J. Baumberg, Single-molecule optomechanics in picocavities,
Science 354, 726 (2016).

\bibitem{park2016} K.-D. Park, E. A. Muller, V. Kravtsov, P. M. Sass, J.
Dreyer, J. M. Atkin, and M. B. Raschke, Variable-temperature tip-enhanced
Raman spectroscopy of single-molecule fluctuations and dynamics, Nano Lett.
16, 479 (2016).

\bibitem{pelton2018} H. Leng, B. Szychowski, M.-C. Daniel, and M. Pelton,
Strong coupling and induced transparency at room temperature with single
quantum dots and gap plasmons, Nat Commun. 9, 4012 (2018).

\bibitem{gross2018} H. Gross, J. M. Hamm, T. Tufarelli, O. Hess, and B.
Hecht, Near-field strong coupling of single quantum dots, Sci. Adv. 2018; 4:
eaar4906.

\bibitem{park2019} K.-D. Park, M. A. May, H. Leng, J. Wang, J. A. Kropp, T.
Gougousi, M. Pelton, M. B. Raschke, Tip-enhanced strong coupling
spectroscopy, imaging, and control of a single quantum emitter, Sci. Adv.
2019;5: eaav5931.


\bibitem{dicke1954} R. H. Dicke, Coherence in spontaneous radiation processes, Phys. Rev. 93, 99 (1954). 

\bibitem{gross1982} M. Gross and S. Haroche, Superradiance: an essay on the theory of collective spontaneous emission, Phys. Rep. 93, 301 (1982). 

\bibitem{schneider2002} S. Schneider, and G. J. Milburn, Entanglement in the steady state of a collective-angular-momentum (Dicke) model, Phys. Rev. A 65, 042107 (2002). 


\bibitem{gonzales2013} A. Gonzalez-Tudela and D. Porras, Mesoscopic entanglement induced by spontaneous emission in solid-state quantum optics, Phys. Rev. Lett. 110, 080502 (2013). 

\bibitem{scully2015} M. O. Scully, Single photon subradiance: quantum control of spontaneous emission and ultrafast readout, Phys. Rev. Lett. 115, 243602 (2015). 

\bibitem{kirton2017}  P. Kirton and J. Keeling, Suppressing and Restoring the Dicke Superradiance Transition by Dephasing and Decay, Phys. Rev. Lett. 118, 123602 (2017). 

\bibitem{wolfe2014} Elie Wolfe and S. F. Yelin, Certifying Separability in Symmetric Mixed States of N Qubits, and Superradiance, Phys. Rev. Lett. 112, 140402 (2014). 

\bibitem{shammah2017} N. Shammah, N. Lambert, F. Nori, and S. De Liberato, Superradiance with local phase-breaking effects, Phys. Rev. A 96, 023863 (2017). 

\bibitem{gegg2018} M. Gegg, A. Carmele, A. Knorr and M. Richter, Superradiant to subradiant phase transition in the open system Dicke model: dark state cascades, New J. Phys. 20,  013006 (2018). 

\bibitem{belyanin1998} A. Belyanin, V.V. Kocharovsky, Vl.V. Kocharovsky, Superradiant generation of femtosecond pulses in  quantum-well heterostructures, Quant. \& Semiclass. Opt. (JEOS Part B) 10, L13-L19 (1998).

\bibitem{cong2016} K. Cong, Q. Zhang, Y. Wang, G. T. Noe II, A. Belyanin, and J. Kono, Dicke superradiance in solids, JOSA B 33, C80 (2016).

\bibitem{temnov2005} V. V. Temnov and U. Woggon, Superradiance and Subradiance in an Inhomogeneously Broadened Ensemble
of Two-Level Systems Coupled to a Low-Q Cavity, Phys. Rev. Lett. 95, 243602 (2005). 

\bibitem{guhne2009} O. Guhne and  G. Toth, Entanglement detection, Phys. Rep. 474, 1 (2009). 



\bibitem{tureci2016} C. Aron, M. Kulkarni, and H. E. T\"{u}reci, Photon-Mediated Interactions: A Scalable Tool to Create
and Sustain Entangled States of N Atoms, Phys. Rev. X 6, 011032 (2016). 



\bibitem{lukin2016} A. Sipahigil, R. E. Evans, D. D. Sukachev, et al., An
integrated diamond nanophotonics platform for quantum-optical networks,
Science 354, 847 (2016).

\bibitem{deppe} T. Yoshie, A. Scherer, J. Hendrickson, G. Khitrova, H. M.
Gibbs, G. Rupper, C. Ell, O. B. Shchekin, and D. G. Deppe, Vacuum Rabi
splitting with a single quantum dot in a photonic crystal nanocavity, Nature
432, 200 (2004).

\bibitem{reithmaier} J. P. Reithmaier, G. Sek, A. Loffler, C. Hofmann, S.
Kuhn, S. Reitzenstein, L. V. Keldysh, V. D. Kulakovskii, T. L. Reinecke, and
A. Forchel, Strong coupling in a single quantum dot - semiconductor microcavity system, Nature 432, 197 (2004).

\bibitem{sercel2019} P. C. Sercel, J. L. Lyons, D. Wickramaratne, R. Vaxenburg,
N. Bernstein, and A. L. Efros, Nano Lett. 19, 4068-4077 (2019). 

\bibitem{fieramosca2019} A. Fieramosca, L. Polimeno, V. Ardizzone, L. De Marco, M. Pugliese, V. Maiorano,
M. De Giorgi, L. Dominici, G. Gigli, D. Gerace, D. Ballarini, D. Sanvitto, Two-dimensional hybrid perovskites sustaining strong
polariton interactions at room temperature, Sci. Adv. 2019;5: eaav9967. 

\bibitem{manj2012} A. Manjavacas, Pe. Nordlander, and F. J. Garcia de Abajo,  Plasmon blockade in nanostructured graphene, ACS Nano 6, 1724 (2012). 

\bibitem{chen2017} Pai-Yen Chen, C. Argyropoulos, M. Farhat and J. S. Gomez-Diaz, Flatland plasmonics and nanophotonics based on
graphene and beyond, Nanophotonics 6, 239 (2017). 






\bibitem{Tokman2016} M.Tokman, Y. Wang, I. Oladyshkin, A. R. Kutayiah, and
A. Belyanin, Laser-driven parametric instability and generation of entangled
photon-plasmon states in graphene, Phys. Rev. B 93, 235422 (2016).

\bibitem{parametric} M. Tokman, M. Erukhimova, Q. Chen, and A. Belyanin, The universal model of strong coupling at the nonlinear  resonance in open cavity-QED systems, Phys. Rev. A 105, 053707 (2022). 






\bibitem{langford2011} N. K. Langford, S. Ramelow, R. Prevedel, W. J. Munro, J. Milburn, and A. Zeilinger, Efficient quantum computing using coherent photon conversion, Nat. 478, 360 (2011).

\bibitem{reitz2022}   M. Reitz, C. Sommer, and C. Genes. Cooperative Quantum Phenomena in Light-Matter Platforms. PRX Quantum 3, 010201 (2022). 



\bibitem{blum} K. Blum, \textit{Density Matrix Theory and Applications}
(Springer, Heidelberg, 2012).










\bibitem{Scully1997} M. O. Scully and M. S. Zubairy, \textit{Quantum Optics}
(Cambridge University Press, Cambridge, 1997).



\bibitem{Gardiner2004} C. Gardiner and P. Zoller, \textit{Quantum Noise}
(Springer-Verlag, Berlin, Heidelberg, 2004).



\bibitem{tokman2018} M. Tokman, Z. Long, S. Al Mutairi,
Y. Wang, M. Belkin, and A. Belyanin, Enhancement of the
spontaneous emission in subwavelength quasi-two-dimensional waveguides and
resonators, Phys. Rev. A 97, 043801 (2018).



 \bibitem{tokman2019} M. Tokman, Z. Long, S. AlMutairi, Y. Wang, V. Vdovin, M. Belkin, and A. Belyanin,  Purcell
enhancement of the parametric down-conversion in two-dimensional nonlinear
materials,  APL Photonics 4, 034403 (2019).

\bibitem{fain} V. M. Fain and Y. I. Khanin, \textit{Quantum Electronics.
Basic Theory} (Cambridge, MA, MIT, 1969).


\bibitem{mollow1975} B. R. Mollow, Pure-state analysis of resonant light scattering: Radiative damping, saturation, and multi-photon effects, Phys. Rev. A 12, 1919-1943 (1975). 




\bibitem{madsen2013} K. H. Madsen and P. Lodahl, Quantitative analysis of
quantum dot dynamics and emission spectra in cavity quantum electrodynamics,
New J. of Phys. 15, 025013 (2013).


\bibitem{Landau1965} L.D. Landau, E.M. Lifshitz, \textit{Statistical
Physics, Part 1} (Pergamon, Oxford, 1965).




\bibitem{may2021}	M. May, T. Jiang, C. Du, K.-D. Park, X. Xu, A. Belyanin, and M. Raschke, Nanocavity clock spectroscopy: resolving competing exciton dynamics in WSe2/MoSe2 heterobilayers, Nano Lett. 21, 522 (2021). 



\bibitem{smythe1968} W.R. Smythe, \textit{Static and Dynamic Electricity}, 3rd ed. (McGraw-Hill, New York, 1968).



\bibitem{griffiths2017} D.J. Griffiths, \textit{Introduction to Electrodynamics}, 4th ed. (Cambridge University Press, Cambridge, 2017).

\bibitem{jackson1999} J.D. Jackson, \textit{Classical Electrodynamics}, 3rd ed. (John Wiley \& Sons, Hoboken, 1999).

\bibitem{landau1984} L.D. Landau, E.M. Lifshitz, \textit{Electrodynamics of Continuous Media}, 2nd ed. (Pergamon, Oxford, 1984).

\bibitem{brand1955} L. Brand, A sequence defined by a difference equation, Am. Math. Mon., 62, 7, 489--492 (1955).



\end{thebibliography}
\end{document}